\documentclass[
  reprint,              
  amsmath,amssymb,      
  aps,                  
  prd,                  
  nofootinbib,          
  superscriptaddress,   
  longbibliography      
]{revtex4-2}
\usepackage[colorlinks=true, linkcolor=blue, citecolor=blue, urlcolor=blue]{hyperref}

\usepackage{graphicx}
\usepackage{lineno}
\usepackage{dcolumn}
\usepackage{bm}
\usepackage{tikz}
\usetikzlibrary{arrows.meta,
                calc, chains,
                quotes,
                positioning,
                shapes.geometric}
                
\usepackage{amsmath} 
\usepackage{listofitems} 
\usepackage[outline]{contour} 
\contourlength{1.4pt}
\usepackage[export]{adjustbox}
\usepackage[normalem]{ulem}

\usepackage{tikz}
\usetikzlibrary{shapes.geometric, arrows, positioning, fit}

\tikzstyle{startstop} = [rectangle, rounded corners, minimum width=3cm, minimum height=1cm, text centered, draw=black, fill=red!30]
\tikzstyle{process} = [rectangle, minimum width=3cm, minimum height=1cm, text centered, draw=black, fill=orange!30]
\tikzstyle{decision} = [diamond, minimum width=3cm, minimum height=1cm, text centered, draw=black, fill=green!30]
\tikzstyle{arrow} = [thick,->,>=stealth]

\tikzset{>=latex} 
\usepackage{xcolor}
\usetikzlibrary{positioning}
\colorlet{myred}{red!80!black}
\colorlet{myblue}{blue!80!black}
\colorlet{mygreen}{green!60!black}
\colorlet{myorange}{orange!70!red!60!black}
\colorlet{mydarkred}{red!30!black}
\colorlet{mydarkblue}{blue!40!black}
\colorlet{mydarkgreen}{green!30!black}
\tikzstyle{node}=[thick,circle,draw=myblue,minimum size=22,inner sep=0.5,outer sep=0.6]
\tikzstyle{node in}=[node, black,draw= black,fill= white]
\tikzstyle{node hidden}=[node,blue!20!black,draw=myblue!30!black,fill=myblue!20]
\tikzstyle{node convol}=[node,orange!20!black,draw=myorange!30!black,fill=myorange!20]
\tikzstyle{node out}=[node,black,draw=black,fill=green!20]
\tikzstyle{connect}=[thick,mydarkblue] 
\tikzstyle{connect arrow}=[-{Latex[length=4,width=3.5]},thick,mydarkblue,shorten <=0.5,shorten >=1]
\tikzset{ 
  node 1/.style={node in},
  node 2/.style={node hidden},
  node 3/.style={node out},
}

\begin{document}

\preprint{APS/123-QED}

\title{Deep Neural Network extraction of Unpolarized Transverse Momentum Distributions}

\author{I. P. Fernando}
\email{ishara@virginia.edu}
\author{D. Keller}%
 \email{dustin@virginia.edu}
\affiliation{%
 Department of Physics, University of Virginia\\
 Charlottesville, Virginia 22904, USA}%

\begin{abstract}
Building on the first-ever application of neural networks in TMD phenomenology—{\it Extraction of the Sivers function with deep neural networks}—we now present a momentum-space, physics-informed deep-learning framework for the direct extraction of unpolarized transverse-momentum–dependent parton distributions (TMDs) from fixed-target Drell–Yan data (E288, E605). Rather than transforming to impact‑parameter space, we remain in $k_\perp$ and embed a normalized integrand $s(x,k_\perp;Q)$ whose auto‑convolution produces the observed $q_T$ spectra. We introduce two phenomenological objects in this data-driven approach: the \textit{transverse structure kernel} $S(q_T,x_1,x_2;Q)$ and the \textit{intrinsic-transverse-momentum profile} $s(x,k_\perp;Q)$. The extraction proceeds in two steps. \emph{Stage I} learns the structure kernel $S(q_T,x_1,x_2;Q_M)$ by regressing the cross‑section with known kinematic prefactors and charge‑weighted PDF combinations factored out; experimental and PDF uncertainties are propagated with Monte Carlo replicas. \emph{Stage II} reconstructs $s(x,k_\perp;Q)$ with an end‑to‑end differentiable $k_\perp$ quadrature layer. Applied to Fermilab cross-section data from experiments E288 and E605, the method reproduces the measured $q_T$ spectra across the $Q_M$ bins and yields $x$‑ and $Q$‑dependent TMDs that broaden with $Q$, with uncertainty bands that consistently propagate experimental, PDF, algorithmic and methodological components. The approach is minimally biased (no factorized \textit{Ansätze} and no $b_T$ transform) and provides a transferable template for polarized TMDs and related QCD inverse problems.
\end{abstract}

\maketitle

\tableofcontents

\section{Introduction}

Transverse-momentum--dependent parton distribution functions (TMDs) provide a fully multidimensional description of hadron structure, encoding the probability densities of partons as functions of both the longitudinal momentum fraction~$x$ and the intrinsic transverse momentum~$k_\perp$~\cite{Ji_1997,XJi2005, boussarie2023tmdhandbook}. 
The TMD factorization formalism~\cite{Ji2004,Ji2005,Vladimirov2022,Chiu2012,Echevarria:2012,Becher2011,Collins2011,Collins1989,Collins1982,Collins1981,Collins1983} enables the separation of perturbatively calculable hard coefficients from nonperturbative TMD distributions in processes differential in transverse momentum. 
Recent comprehensive extractions of unpolarized TMDs can be found in Refs.~\cite{Anselmino2014,Bacchetta2017,Bertone2019,Scimemi2019,Vladimirov2019,Bury2022,Moos2024,MAP2022,Boglione2022,Barry2023,Boglione2023,MAP2024,Aslan_Ted_2025,Bacchetta2025,Moos_2025,Barry2025, Cerutti_2023, rossi2025}, polarized TMDs in Refs. \cite{Bury2021a,Bury2021b,Horstmann2023,Yang2025}), and lattice QCD calculations \cite{He_2024}.
The unpolarized TMD, $f_{1,q/N}(x,k_\perp;\mu,\zeta)$, defines the baseline partonic distribution from which all spin-dependent TMDs are normalized and interpreted. 
Moreover, $f_{1,q/N}$ is essential for the quantitative description of semi-inclusive deep-inelastic scattering (SIDIS), Drell--Yan (DY), and vector-boson production across both fixed-target and collider kinematics~\cite{EIC_Boer_2011,EIC_Accardi_2016,EIC_Abdul_2022}, where $\mu$ denotes the renormalization scale and $\zeta$ the rapidity evolution scale.

Historically, extractions of unpolarized TMDs have been carried out in both transverse-momentum space ($k_\perp$) and impact-parameter space ($b_T$). The modern $b_T$-space formalism, established within the Collins--Soper--Sterman (CSS) framework~\cite{Collins2011,CSS1985}, transforms the transverse convolutions into simple products, thereby rendering TMD evolution more tractable and compact. This has made the $b_T$ representation the current standard approach for contemporary global analyses. High-precision extractions in $b_T$ space have now achieved NNLL--N$^3$LL accuracy, providing excellent agreement with extensive experimental cross-section data sets~\cite{Scimemi2018,Scimemi2019,Bacchetta2017,MAP2022,MAP2024,MAP2025} as well as transverse single spin asymmetries (Sivers) \cite{Bury2021a,Bury2021b,Echevarria:2014SiversEvol,Bacchetta:2022,Kang:2016GlobalSivers} and (Collins and Transversity) \cite{Anselmino:2007CollinsSIDIS,Anselmino:2013Transversity,Kang:2015CollinsEvol,Kang:2020TransversityGlobal}. However, many of the competing results for $f_{1,q/N}$ are inconsistent with each other, highlighting the need for a more universal and robust extraction methodology.

Nevertheless, the transverse momentum itself is what mostly closely connects to experimental observables. Performing extractions natively in $k_\perp$ space offers the most transparent link to measured spectra and can therefore exploit the full information content of the data, especially when combined with deep neural networks (DNNs) to capture nontrivial correlations and nonlinear dependencies. However, such an approach poses significant analytical and computational challenges due to the inherently two-dimensional angular convolutions in transverse momentum~\cite{Ji2005,Vogelsang_2005,Schweitzer2010}.

Early phenomenological studies performed directly in $k_\perp$ space typically adopted simple, non-evolving Gaussian parametrizations for the transverse-momentum distribution, with widths assumed to be independent of $x$ (and, in fragmentation, of $z$)~\cite{Anselmino2005,Anselmino2013}. 
These models provided an important first baseline for describing low-$P_T$ spectra and spin-averaged observables, but they inherently imposed a factorized form between the longitudinal and transverse degrees of freedom. 
Incorporating QCD evolution consistently within this scheme necessitated reformulating the description in $b_T$ space, where the Collins--Soper--Sterman (CSS) evolution could be implemented systematically~\cite{Anselmino2013}. 

Accumulating experimental evidence has since demonstrated that such a fully factorized picture is inadequate: SIDIS multiplicities reveal a $P_T$ dependence that varies with both $x$ (at fixed $z$) and $z$ (at fixed $x$), indicating that $\langle k_T^2\rangle$ carries a nontrivial $x$ dependence, while $\langle P_T^2\rangle$ in fragmentation depends explicitly on $z$~\cite{Signori2013,COMPASS2018}. 

Modern analyses have therefore transitioned to $b_T$-space frameworks with flexible nonperturbative sectors, often incorporating flavor and kinematic dependence in the transverse widths. 
These fits typically achieve NLL to N$^3$LL precision and describe both SIDIS and Drell--Yan data—including fixed-target measurements from E288 and E605~\cite{Ito1981,Moreno1991,Bacchetta2017,Bacchetta2020,MAP2022,MAP2024}—with excellent accuracy. 


Following our initial introduction of deep neural-network techniques for extracting TMDs (Sivers Function) \cite{Fernando2023}, and during the ongoing development of our $ k_T $-space analysis of the Unpolarized TMD, Bacchetta et al. \cite{Bacchetta2025} reported a simple neural-network extraction of unpolarized TMDs in impact-parameter ($  b_T  $) space. In parallel, the hadron-structure–oriented formulation introduced in Ref.~\cite{Aslan_Ted_2025} unifies low- and high-energy Drell--Yan and $Z^0$ production within a consistent theoretical framework, offering new insights into the interplay between perturbative and nonperturbative transverse dynamics. 
Even more recent results~\cite{Moos_2025} have advanced this frontier with a global extraction at N$^4$LL accuracy, incorporating flavor-dependent TMDs and fragmentation functions together with the first application of transverse-momentum moments (TMMs). We note that the level of perturbative accuracy in ~\cite{Moos_2025} maybe lower than $N^{4LL}$, due to the requirements of OPE matching to collinear fragmentation functions at $N^{3LO}$,
which are not yet available.


At the core of the present work, artificial intelligence (AI) provides a uniquely powerful framework for TMD extraction. 
Flexible neural architectures can capture nonparametric, multiscale correlations across $(x, Q, k_\perp)$ while embedding physics constraints as inductive biases and propagating experimental uncertainties through replica ensembles. 
Building upon this principle, our previous study performed the first-ever neural-network-based extraction of a TMD—the Sivers function—by globally fitting SIDIS data ~\cite{Fernando2023}. 
That work introduced the concept of implicit inclusion and multiplicative compositionality (or functional multiplicativity), demonstrating that even nominally factorized representations can yield minimally biased TMD information when the neural component possesses sufficient expressive capacity. 
This established a data-driven, physics-informed paradigm for TMD phenomenology and directly motivates the present analysis, which extends the framework by embedding the neural network within the $k_\perp$ convolution integral itself, enabling fully end-to-end differentiable training and direct optimization against experimental observables.

In this work, we pursue a complementary approach to traditional $b_T$-based extractions by performing the entire analysis directly in momentum space and exploring how modern computational methods and AI can be optimally leveraged for nonperturbative QCD inference. 
The prevailing preference for the $b_T$ representation in global analyses arises primarily from the mathematical convenience of the inverse problem: the Fourier–Bessel transform converts transverse-momentum convolutions into products, enabling a compact implementation of CSS evolution and resummation~\cite{CSS1985,Collins2011,Bracewell1999}. 
Moreover, lattice-QCD calculations of the CSS evolution kernel are naturally formulated in impact-parameter space~\cite{Shanahan_2020}, providing valuable first-principles input on the nonperturbative $b_T$ dependence. 
Nevertheless, analyzing the transverse structure directly in $k_\perp$ space offers distinct advantages for data-driven approaches: it preserves a one-to-one correspondence with the measured spectra, allows experimental uncertainties and kinematic effects to be incorporated without intermediate transformations, and facilitates the inclusion of differentiable, AI-based forward models. 
Accordingly, both $b_T$ and $k_\perp$ representations remain essential and mutually constraining perspectives on TMD dynamics.

\vspace{0.5em}
\noindent
Our strategy is to render the $k_\perp$ inverse problem tractable by numerical means rather than by changing representation. 
We embed a DNN directly as the integrand of the $k_\perp$-space convolution and train it on measured cross-sections in the small-$q_T$ region where TMD factorization holds, thus eliminating the need to transform to $b_T$ for fitting and then back to $k_\perp$ for interpretation. 
Concretely, we represent the unpolarized TMD as
\[
  f_{1,q/N}(x,k_\perp;Q,Q^2)
  \;=\;
  f_{1,q}(x;Q^2)\, s(x,k_\perp;Q)\,,
\]
where we introduce the {\it intrinsic-transverse-momentum profile} $s(x,k_\perp;Q)$: a single nonparametric profile learned from data. 
Unlike factorized \textit{ansätze}, this joint representation allows the fit to capture correlations among $x$, $Q$, and $k_\perp$ supported by the data, while the per-point normalization condition $\int d^2k_\perp\, s(x,k_\perp;Q)=1$ for all $(x,Q)$ fixes the overall scaling. 
This normalization preserves the PDF--TMD matching relation~\cite{Aslan_Ted_2025},
\begin{align}
\label{eq:pdf-matching}
    f_{1,q}(x;Q^2) &= \int d^2{\bf k}\, f_{1,q}(x,k_\perp;Q^2)\,,
\end{align}
ensuring consistency with collinear PDFs.

The forward map from network parameters to observables remains fully differentiable through a quadrature layer that performs the two-dimensional angular convolution \emph{within the training loop}~\cite{Baydin2018AD}. 
We do not impose an explicit resummed Sudakov kernel or a fixed-order $Y$-term
(cf. the CSS $W{+}Y$ decomposition~\cite{CSS1985,Collins2011}); instead, we set $\mu=\sqrt{\zeta}=Q$, take the Drell--Yan hard factor at tree level, and allow the joint profile $s(x,k_\perp;Q)$ to absorb the $Q$ dependence directly from the data in the small-$q_T$ region. 
Here $Q$ (or equivalently $Q_M$) denotes the dilepton invariant mass and serves as the hard scale in Drell--Yan, analogous to the photon virtuality $Q^2$ in SIDIS. 
Because no perturbative resummation of cusp or non-cusp anomalous dimensions is applied, it is not meaningful to assign an $\text{N}^k\text{LL}$ accuracy label to this extraction. 
Our results should therefore be characterized as a data-driven momentum-space analysis with a tree-level hard factor and no explicit logarithmic resummation, complementary to $b_T$-space fits where $\text{N}^k\text{LL}$ accuracy is defined by the perturbative content of the Sudakov kernel and matching~\cite{Collins2011,CSS1985,Echevarria:2012,Scimemi2019,Bacchetta2017,Landry2003,Konychev:2006}. 
This distinction is crucial: whereas $b_T$-space fits infer the $Q$ dependence from perturbative inputs, our momentum-space framework learns it directly from the data, reducing bias not only by avoiding predefined functional \textit{ansätze} for the $k_\perp$ dependence but also by dispensing with external priors from resummation schemes and the intermediate $b_T$ mapping. 
The joint profile $s(x,k_\perp;Q)$ is thus determined directly from experiment, with no structural constraints beyond PDF matching, providing a minimally biased, data-driven approach to TMD extraction.

\vspace{0.5em}
\noindent
The analysis follows standard fixed-target Drell--Yan practice. 
We reconstruct $(x_1,x_2)$ from $(Q_M,y)$ at leading order in $q_T/Q_M$ as
$x_{a,b}=(Q_M/\sqrt{s})\,e^{\pm y}$~\cite{Collins2011}, 
restrict the fit to the TMD domain $q_T\ll Q_M$ with the condition $q_T < 0.2 ~ Q_M$ where the $W{+}Y$ separation is valid~\cite{CSS1985,Collins2011}, 
veto the $\Upsilon$ resonance region in E288/E605~\cite{Ito1981,Moreno1991}, 
choose $\mu=\sqrt{\zeta}=Q$ as a canonical scale~\cite{Collins2011}, 
and neglect the $Y$-term in fits limited to small $q_T$~\cite{CSS1985}. 
Uncertainties are propagated through Monte-Carlo replica ensembles constructed from the reported pointwise errors of E288/E605 (full covariances are not available)~\cite{Ito1981,Moreno1991}, 
and collinear-PDF uncertainties are incorporated by sampling NNPDF4.0 replicas accessed through LHAPDF~\cite{Ball2022,Buckley2015LHAPDF}. 
Within this controlled kinematic window, the method provides a direct, minimally biased extraction in $k_\perp$ space, complementary to global $b_T$-space analyses that emphasize resummed evolution and matching~\cite{Collins2011,Scimemi2019,Bacchetta2017}. 
We note that the differentiable-programming framework employed here can, with minimal modification, be adapted to a $b_T$ representation~\cite{Baydin2018AD}.

\vspace{0.5em}
\noindent
Our results demonstrate the feasibility of solving the inverse problem entirely in momentum space, without factorizing longitudinal and transverse dynamics or mapping between $k_\perp$ and $b_T$. 
In this sense, the present work establishes a data-driven alternative path to unpolarized TMDs that remains directly connected to the measured observables. 
Methodologically, we employ an end-to-end differentiable inverse-convolution integration in $k_\perp$ that learns correlated $x$–$Q$–$k_\perp$ structure directly from the cross-sections, aligning with modern learned-operator approaches to inverse problems~\cite{Arridge2019Inverse,Adler2018PrimalDual}.

The remainder of the paper is organized as follows. 
Section~\ref{sec2} outlines the DY formalism in $k_\perp$ space and defines the joint, normalized representation $s(x,k_\perp;Q)$ used throughout. 
Section~\ref{sec:methodology} details the momentum–space extraction pipeline and describes the two-stage methodology: 
Stage I, the cross-section–level learning of the {\it{transverse structure-kernel}} $S(q_T;x_1,x_2,Q_M)$, and 
Stage II, the physics-informed reconstruction of the {\it{intrinsic-transverse-momentum profile}} $s(x,k_\perp;Q)$ with its recursive uncertainty propagation scheme. 
Section~\ref{sec:closure} presents closure tests with pseudo-data, validating the end-to-end inversion and quantifying methodological and algorithmic uncertainties. 
Section~\ref{sec5:fits} applies the full procedure to the E288 and E605 fixed-target data. 
Sections~\ref{sec:stage1-Sfit} and \ref{sec:inverse-kperp-film} describe, respectively, the DNN implementation of the $S(q_T;x_1,x_2,Q_M)$ fit and the differentiable $k_\perp$ inverse convolution that reconstructs $s(x,k_\perp;Q)$. 
Section~\ref{sec:results} presents the extracted TMDs, their scale evolution, and the uncertainty decomposition, while Section~\ref{sec:conclusion} summarizes the conclusions and outlook.

\section{\label{sec2}Formalism}

We work in the TMD factorization regime of the DY process, i.e.\ at small transverse momentum $q_T\equiv|{\bf q}_T|\ll Q_M$ (with $Q_M$ the dilepton invariant mass), where the cross-section factorizes into hard, soft, and (TMD) collinear functions~\cite{Ji_1997,Ji2004,XJi2005,Collins2011,Bacchetta2008}. The triple-differential cross-section can be written as
\begin{widetext}
\begin{align}
    \frac{d\sigma}{dq_T\, dQ_M\, dy}
    &= \frac{16\pi^2 \alpha^2}{9\,Q_M^3}\; q_T \; F^{1}_{UU}(x_1,x_2,|q_T|,Q),
    \label{eq:master_xsec}
\end{align}
where the unpolarized structure function $F^{1}_{UU}$ is
\begin{align}
    F^{1}_{UU}
    = x_1 x_2 \sum_{q} e_q^2 \,\mathcal{H}^{\mathrm{DY}}(Q,\mu)
    & \left( \int d^{2}{\bf k}_{1T}\, d^{2}{\bf k}_{2T}\,
    f_{q/1}(x_1,{\bf k}_{1T};\mu,\zeta_1)\,
    f_{\bar q/2}(x_2,{\bf k}_{2T};\mu,\zeta_2)\, \right. \nonumber \\
    & \left. \times \delta^{(2)}\!\left({\bf q}_T-{\bf k}_{1T}-{\bf k}_{2T}\right) \right)
    + (q \leftrightarrow \bar{q}). 
    \label{eq:FUU_conv}
\end{align}
\end{widetext}
Here $y$ is the dilepton rapidity, and $(x_1,x_2)$ are the partonic momentum fractions satisfying, at leading order in $q_T/Q_M$,
\begin{align}
x_1=\frac{Q_M}{\sqrt{s}}\,e^{+y},~
x_2=\frac{Q_M}{\sqrt{s}}\,e^{-y},~
x_1 x_2=\frac{Q_M^2}{s}\, .
\label{eq:xa_xb}
\end{align}
The scales $(\mu,\zeta_{1,2})$ denote the renormalization and Collins--Soper (CS) scales associated with TMD evolution~\cite{Collins2011,Scimemi2019}. The hard factor $\mathcal{H}^{\mathrm{DY}}(Q,\mu)=1+\mathcal{O}(\alpha_s)$ contains the perturbative matching coefficients for the partonic subprocess.

For datasets reported in terms of the Feynman variable $x_F$ (notably E605),
we convert to rapidity using the leading-order kinematic relations valid in the
small-$q_T/Q$ limit:
\begin{align}
x_F = x_1 - x_2 = \frac{2Q}{\sqrt{s}}\sinh y,
\qquad
y = \sinh^{-1}\!\left(\frac{x_F\sqrt{s}}{2Q}\right),
\label{eq:xf_to_y}
\end{align}
together with $x_1 x_2 = Q^2/s$.
In practice, we use Eq.~\eqref{eq:xf_to_y} to populate $(y,x_1,x_2)$ from the
published $(Q,x_F)$ bin centers before evaluating the forward map. Thus, for E605 cross-section we used,
\begin{align}
E\;\frac{d\sigma^{\mathrm{DY}}}{d^3q}
=\frac{\bar{Q}\cosh{\bar{y}}}{\pi q_T \sqrt{s}}\,\frac{d\sigma^{\mathrm{DY}}}{dq_T\,dy}\, ,
\end{align}
\label{cs-E605}
where $\bar{Q},\bar{y}(\bar{x}_F,\bar{Q})$ are bin-averaged values.

To keep the data-driven inference aligned with the measured $q_T$ spectra and to minimize
representation-induced modeling choices associated with transforming to $b_T$ and back,
we perform the extraction natively in momentum space and adopt a single, joint profile for the
\emph{unpolarized} TMD,
\begin{align}
    f_{q/h}(x,k_\perp;Q,Q^2)
    \;=\;
    f_{q/h}(x;Q^2)\; s_{h}(x,k_\perp;Q)\,,
    \label{eq:joint_ansatz}
\end{align}
where $h\in\{a,b\}$ labels the incoming hadron and
$s_{h}(x,k_\perp;Q)$ is a nonparametric profile that captures correlations among
$x$, $k_\perp$, and $Q$.  Equation~(\ref{eq:joint_ansatz}) is applied only in the TMD region; throughout this work we restrict to
$q_T<0.2\,Q$ and do not include a $Y$ term or fixed-order matching by design, so the extracted
profiles should be interpreted as data-drive momentum-space TMDs within this domain.
The per-\((x,Q)\)-point normalization
\begin{align}
    \int d^2 k_\perp \; s_{h}(x,k_\perp;Q)
    \;=\; 1,
    \qquad \forall\,(x,Q)\in\mathbb{R}\times\mathbb{R}_{>0},
    \label{eq:perpoint_norm}
\end{align}
ensures consistency with the collinear PDFs and fixes the overall normalization.

In Eq.~\eqref{eq:joint_ansatz} we employ a single transverse-momentum profile
$s_h(x,k_\perp;Q)$ that is shared among quark flavors within a given hadron $h$.
This should be understood as a controlled modeling choice motivated by the limited
flavor sensitivity of the fixed-target Drell--Yan data considered here (E288 and E605),
which predominantly constrain PDF-weighted quark--antiquark combinations rather than
independent flavor-resolved intrinsic transverse-momentum distributions.
Even with a common profile $s_h$, the assembled TMDs
$f_{q/h}(x,k_\perp;Q,Q^2)=f_{q/h}(x;Q^2)\,s_h(x,k_\perp;Q)$ retain nontrivial flavor
dependence through the collinear PDFs.

Modern global extractions that incorporate broader datasets
(e.g.\ collider Drell--Yan/$Z$ measurements and, in many cases, SIDIS)
can resolve additional flavor dependence in the transverse sector and therefore
employ flavor-dependent $k_\perp$ profiles.
Importantly, the present inverse-problem framework is modular and can be generalized
straightforwardly by promoting $s_h\to s_{q/h}$ (equivalently
$\lambda_h\to\lambda_{q/h}$) and conditioning the integrand network on a discrete
flavor label.
We defer such extensions to future global analyses where the corresponding degrees
of freedom are better constrained by data.

With Eq.~\eqref{eq:joint_ansatz}, the $k_\perp$-space convolution in Eq.~\eqref{eq:FUU_conv} reduces to a two-point angular auto-convolution of the profiles $s$:
\begin{align}
 & S^{(ab)}(q_T; x_1,x_2,Q)
 \nonumber \\
& \equiv \int_0^\infty\! dk_\perp\, k_\perp
   \int_0^{2\pi}\! d\phi\;
   s_{a}(x_1,k_\perp,Q)\nonumber\\
&\qquad\times
   s_{b}\!\Bigl(
      x_2,
      \sqrt{\,q_T^2+k_\perp^2-2 q_T k_\perp\cos\phi\,},
      Q
   \Bigr).
\label{eq:S_def_joint}
\end{align} 
and the structure function becomes
\begin{widetext}
\begin{align}
\begin{split}
F^{1}_{UU}
= \mathcal{H}^{\mathrm{DY}}(Q,\mu)\,
  \sum_{q} e_q^2 \Bigl[
&\; x_1 f_{q/a}(x_1;Q^2)\,x_2 f_{\bar q/b}(x_2;Q^2)\,
   S^{(ab)}(q_T;x_1,x_2,Q)\\
&\; +\; x_1 f_{\bar q/a}(x_1;Q^2)\,x_2 f_{q/b}(x_2;Q^2)\,
   S^{(ab)}(q_T;x_1,x_2,Q)
\Bigr].
\end{split}
\label{eq:FUU_joint}
\end{align}
\end{widetext}
For $pp$ scattering with flavor-symmetric transverse profiles, one may have $S^{(ba)}=S^{(ab)}$; for proton--nucleus kinematics, distinct beam/target profiles $s_{a}$ and $s_{b}$ can be retained without further assumptions, which we will address in future work.

For comparison with legacy presentations of fixed-target data, we use
\begin{align}
E\;\frac{d\sigma^{\mathrm{DY}}}{d^3q}
=\frac{1}{2\pi q_T}\,\frac{d\sigma^{\mathrm{DY}}}{dq_T\,dy}\, .
\label{eq:inv}
\end{align}
We set $\mu=\sqrt{\zeta}=Q$ and restrict to the small-$q_T$ region where TMD factorization applies, neglecting the $Y$-term that restores the fixed-order tail at $q_T\sim Q_M$~\cite{Collins2011,Bacchetta2008}. Collinear inputs $f_{q/h}(x;Q^2)$ are taken from modern global fits (e.g.\ NNPDF4.0) via LHAPDF~\cite{Ball2022,Buckley2015LHAPDF}.

In Eq.~\eqref{eq:S_def_joint} the profiles $s_{h}(x,k_\perp;Q)$ are represented by a physics-informed deep neural network (DNN) and embedded directly as the \emph{integrand} of the transverse-momentum convolution. A differentiable quadrature layer carries out the radial and angular integrations, so that the forward map from network parameters to the observable cross-sections is smooth and trainable by gradient methods. This design allows us to learn nontrivial $x$--$k_\perp$--$Q$ correlations without imposing factorized functional forms. When training on experimental data (E288/E605 \cite{Ito1981,Moreno1991}), the measured cross-sections are provided in finite $(Q_M,y,q_T)$ bins; we evaluate the theory at the bin centers, include the Jacobian in Eq.~\eqref{eq:inv}, and have verified that bin-width effects are negligible at the present level of precision. Although the $b_T$ representation remains mathematically convenient---the analogue of Eq.~\eqref{eq:S_def_joint} becomes a product after a Bessel--Fourier transform---our analysis stays entirely in $k_\perp$ space and does not rely on mapping to $b_T$ for fitting. This keeps the extraction close to the measured $q_T$ spectra while retaining the ability to cross-check against $b_T$-space results when needed.

\section{\label{sec:methodology}Methodology}

Our extraction proceeds in momentum space and is organized in two stages. In Stage~I we learn the \textit{transverse structure kernel} \(S(q_T;x_1,x_2,Q_M)\) directly from measured DY cross-sections after factoring out known kinematic prefactors and collinear PDFs. In Stage~II we solve the inverse problem in \(k_\perp\) by training a physics‑informed integrand \(s(x,k_\perp,Q)\) whose auto‑convolution reproduces the learned \(S\) with per‑\((x,Q)\) normalization and regularity constraints. The unpolarized TMD is finally assembled as
\begin{align}
  f_{1,q/N}(x,k_\perp;Q,Q^2)=f_{1,q}(x;Q^2)\,s(x,k_\perp;Q),
\end{align}
with \(\int d^2k_\perp\,s(x,k_\perp;Q)=1\) enforced for every \((x,Q)\) in the fit support, where the flavor-dependence is carried by the collinear-PDFs.

We use fixed‑target DY cross-sections from E288 (200, 300, 400~GeV) \cite{Ito1981} and E605 (800~GeV) \cite{Moreno1991}. We test our output with the standard small‑\(q_T\) cuts for TMD factorization, reconstruct \((x_1,x_2)\) from \((Q_M,y)\) at leading order in \(q_T/Q_M\), exclude the \(\Upsilon\) region by vetoing \(9~\mathrm{GeV}\!<\!Q_M\!<\!11~\mathrm{GeV}\), and treat measured cross-sections in finite bins. For each \(Q_M\) bin, the width $\Delta Q_M$ is taken from the data, and we approximate the bin integral analytically via

\begin{align}
  \int_{Q_M-\frac{\Delta Q_M}{2}}^{Q_M+\frac{\Delta Q_M}{2}}\!\!\!\!\frac{dQ'}{Q'^3}
  = \frac{1}{2}\!\left(\frac{1}{(Q_M-\tfrac{\Delta Q_M}{2})^2}-\frac{1}{(Q_M+\tfrac{\Delta Q_M}{2})^2}\right)
  \label{Eq11}
\end{align}
which enters the rescaling from the triple‑differential form in Eq.~\eqref{eq:master_xsec} to our learning target (see Section \ref{subsec:3-A}). Sample‑dependent weights are applied to balance under‑represented kinematic slices; these are fixed once at the data‑preparation stage and used consistently throughout training. Equation~(\ref{Eq11}) treats the explicit tree-level prefactor $\propto Q^{-3}$ exactly over the
finite mass bin, while evaluating the remaining (smooth) factors at the bin center $Q_M$;
this is a midpoint approximation whose leading correction scales as $\mathcal{O}(\Delta^2)$
for symmetric bins. It's also possible to evaluate the full bin
integral numerically by
adding a low-order quadrature over $Q'$ inside the differentiable forward map.

\subsection{Stage I: cross‑section–level learning of \(S(q_T;x_1,x_2,Q_M)\)}
\label{subsec:3-A}
We build a supervised model that maps $(q_T,x_1,x_2,Q_M)\mapsto S(q_T;x_1,x_2,Q_M)$ and is trained against the measured cross-section after removing known prefactors and isolating the PDF combinations. Concretely, we write 
Eq.~\eqref{eq:FUU_joint} as
\begin{widetext}
\begin{align}
  \frac{d\sigma}{dq_T\, dQ_M\, dy}\;=\frac{(4\pi\alpha)^2}{9\cdot 2\pi}\;\left[\int_{Q\text{-bin}}\!\!\frac{dQ'}{Q'^3}\right] 
  \sum_q e_q^2 & \left[x_1 f_{q/a}(x_1;Q_M^2)\,x_2 f_{\bar q/b}(x_2;Q_M^2) S(q_T;x_1,x_2,Q_M) \right. \nonumber \\
  & \left. +x_1 f_{\bar q/a}(x_1;Q_M^2)\,x_2 f_{q/b}(x_2;Q_M^2)\,S(q_T;x_1,x_2,Q_M)\right],
\label{eq:cs-setup}
\end{align}
\end{widetext} 
where the prefactor$\frac{(4\pi \alpha)^2}{18\pi}$
converts to the experimental units used by E288/E605. The per-point training target is then
\begin{align}
  S_{\rm calc}(q_T;x_1,x_2,Q_M)
  &= \frac{\left(\frac{d\sigma}{dq_T\, dQ_M\, dy}\right)}{\tfrac{(4\pi\alpha)^2}{18\pi}} \nonumber\\
  &\quad\times 
     \left[\int_{Q\text{-bin}} \frac{dQ'}{Q'^{3}}\right]^{-1}.
\label{eq:Scalc}
\end{align}
while the inputs carry two auxiliary channels with precomputed PDF weights for the \(q\bar q\) and \(\bar q q\) orderings. The network outputs the PDF‑weighted sum of \(S\) contributions and is trained with a weighted mean‑squared loss using the experimental per‑point weights. Progressive training with an early‑stopping rule is employed to stabilize optimization. After training a model ensemble over cross‑section replicas and PDF replicas, we project the network(s) onto a fine grid in \((x_1,x_2,q_T,Q_M)\) and explicitly evaluate both orderings by swapping \((x_1,x_2)\) in the input. We impose $pp$ beams-targets exchange symmetry such that $S{q\bar{q}=S{\bar{q}q}}$, intentionally neglecting nuclear effects from the target.
For proton–proton kinematics the kernel may be symmetric under the exchange
$(x_1,x_2)\!\leftrightarrow\!(x_2,x_1)$, i.e.
\(
S(q_T;x_1,x_2,Q)=S(q_T;x_2,x_1,Q).
\)
Accordingly, in Stage~I we impose this symmetry and define the symmetrized predictor
\begin{align}
&S_{\rm sym}(q_T;x_1,x_2,Q) = \nonumber\\
&\tfrac{1}{2}\Big[S_{\theta}(q_T;x_1,x_2,Q)
   + S_{\theta}(q_T;x_2,x_1,Q)\Big] .
\label{eq:Ssym}
\end{align}
which is used both in the training loss and when projecting to the $(x_1,x_2,q_T,Q)$ grid.%
\footnote{Equivalently, one can enforce symmetry by (i) sorting $(x_1,x_2)$ at the input,
(ii) using symmetric invariants, e.g.\ $u=x_1{+}x_2$, $v=x_1x_2$, or
(iii) adding a small penalty $\lambda_{\rm sym}\!\left|S_\theta(q_T;x_1,x_2,Q)-S_\theta(q_T;x_2,x_1,Q)\right|^2$
to the loss.}
For each replica $r$ we compute $S_{\rm sym}^{(r)}$ on the grid; the Stage~I label is the
replica mean and band,
\(
\overline{S}_{\rm sym}=\langle S_{\rm sym}^{(r)}\rangle_r
\)
with $1\sigma$ dispersion from the ensemble. We use $\overline{S}$ (for $\overline{S}_{\rm sym}$) in the rest of the article for convenience. These
$\overline{S}$ labels are then used in Stage~II to reconstruct $s(x,k_\perp,Q)$.
For reference, the Stage-I network starts with 4 input nodes for the kinematics $(q_T,x_1,x_2,Q_M)$ and intermediate layers with widths $(128,128,128,64)$, totaling
$N_{\rm par}=41,985$ trainable parameters including the output node, with the activation choices: `swish' (for intermediate layers) and `softplus' for output layer.
For proton–nucleus kinematics this exchange symmetry would \emph{not} be imposed as in Eq.~\eqref{eq:Ssym} and the two
orderings would be kept distinct.  This type of analysis will be addressed in future work.\\

\subsection{Stage II: physics‑informed reconstruction of \(s(x,k_\perp,Q)\) in \(k_\perp\) space}
The Stage~II network represents the normalized profile \(s(x,k,Q)\) and is trained directly against the Stage~I labels \(\overline{S}\) through the momentum‑space auto‑convolution

\begin{widetext}
\begin{align}
S(q_T;x_1,x_2,Q)\;=\;\int_0^{2\pi}\!d\phi\int_0^\infty\!dk\,k\;
s(x_1,k;Q)\;s\!\left(x_2,\sqrt{q_T^2+k^2-2q_Tk\cos\phi};Q\right).
\label{eq:convolution_stage2}
\end{align}
\end{widetext}
To avoid ad hoc functional forms, we parametrize the $s(x,k,Q)$ logarithmic slope using a compact Feedforward DNN with \(\lambda(x,k,Q)\ge0\) and set
\begin{align}
  s(x,k,Q) &= \exp\!\left[-\int_0^k \lambda(x,\kappa,Q)\,d\kappa\right], \nonumber\\
  s(x,0,Q) &= 1 \, ,
\label{eq:s_definition}
\end{align}
so monotonicity and integrability follow by construction (see Sec.~\ref{sec:lambda-net} for the details on the DNN architecture for $\lambda(x,k,Q_M)$). The per‑\((x,Q)\) normalization \(\int d^2{\bf k}\,s(x,k_\perp;Q)=1\) is imposed softly on a high‑order Gauss–Legendre grid with a cosine taper that switches on at \(k\simeq0.85\,k_{\max}\) and a tail mask beyond \(0.9\,k_{\max}\). The forward operator evaluates Eq.~\eqref{eq:convolution_stage2} with \(N_\phi=64\) angular points and an adaptive radial grid of \(R_{\rm nodes}=4096\) nodes up to \(k_{\max}=5\), using a trapezoidal rule in \(k\) and precomputed targets for physics anchors (see below). The full map from network parameters to \(S\) is differentiable, enabling end‑to‑end gradient training.

The map from the normalized profile $s(x,k;Q)$ to the projected observable
$\overline{S}(q_T;x_1,x_2,Q)$ is an integral (auto--convolution) operator and is therefore
smoothing. As a result, the inverse reconstruction is not uniformly well conditioned in the
radial variable $k$: variations of $s$ in some $k$ regions
can induce only weak changes in the resolved $q_T$ spectra.
To ensure numerical stability of the inversion, we retain the mass constraint by definition
and, during model development, we introduce auxiliary soft regularizers to suppress
unphysical oscillations in weakly constrained directions.
These auxiliary terms are used only for prototyping and are removed for the final production
extraction once the architecture and hyperparameters have been fixed by recursive closure tests. Further details are provided in Section \ref{sec:inverse-kperp-film}.

\subsection{\label{sec3-3}Uncertainties and the recursive extraction scheme}

Our uncertainty quantification adheres to a robust recursive framework \cite{Fernando2023}, which is designed to stabilize extraction during error quantification through the process of propagation of experimental errors. This approach systematically distinguishes between precision and accuracy, while also quantifying contributions from algorithmic and methodological sources. The approach iterates between pseudo-data closure and real data fitting until the pseudo-data reproduce the real cross-sections within experimental errors. In each outer iteration \(r=0,1,2,\dots\) we:
\begin{enumerate}
    \item Generate pseudo-data from a current generator \( G^{(r)} \) over the full \( (x_1, x_2, q_T, Q_M) \) support and train the Stage~I model ensemble to closure, then project and average to produce \( \overline{S}^{(r)} \);
    \item Reconstruct \( s^{(r)}(x, k, Q) \) by Stage~II training against \( \overline{S}^{(r)} \), propagate to observables, and compare with the experimental cross-sections;
    \item Update the pseudo-data generator and its nuisance settings so that the next round better matches the real data within the quoted uncertainties.
\end{enumerate}

The iteration stops when pseudo‑data and real‑data predictions agree (bin‑wise) within the experimental errors. The \emph{accuracy} of the method is then assessed by the kinematically resolved residual between the known generator truth and the final extraction on the matched pseudo‑data,
\begin{align}
\label{delta_acc}
  \Delta_{\rm acc}(x,k,Q)\;=\;s_{\rm extr}^{(r_\star)}(x,k,Q)\;-\;s_{\rm true}^{(G^{(r_\star)})}(x,k,Q),
\end{align}
with \(r_\star\) the last iteration. The \emph{precision} is quantified by the spread of the replica ensemble at the level of \(S\) (Stage~I) and propagated to \(s\) (Stage~II), while the \emph{methodological} component collects controlled variations described next. This recursive strategy reduces variance, mitigates initialization sensitivity, and provides a transparent, kinematics‑dependent separation of precision and accuracy; it follows the schema outlined in our earlier work and adapted here to the unpolarized case.

Because full experimental covariance matrices are not available for the E288/E605 sets, we propagate experimental uncertainties by Monte Carlo replicas of the cross-sections in Stage~I and by PDF replicas in the collinear inputs. The Stage~I ensemble induces an ensemble of \(S\) grids whose mean and \(1\sigma\) dispersion define the labels and their “precision” bands for Stage~II. On top of this statistical component the following are also strictly quantified and minimized:

(i) Algorithmic error from network initialization, optimizer settings, and training schedules (tracked by repeating trainings with different seeds and schedules, quantified without experimental errors included);

(ii) Integration‑scheme error from changing the Stage~II evaluator (direct angular–radial quadrature vs.\ auto‑convolution) and from tightening/loosening \(N_\phi\), \(R_{\rm nodes}\), and \(k_{\max}\). There is a separate component of methodological error found from the residual of $s^{(r)}(x,k,Q)$ and a test function $s^{(r_*)}_{true}(x,k,Q)$) (the mean model of the final iteration);

(iii) Experimental input error from the cross-sections (already included via the Stage~I replica ensemble);

(iv) Collinear‑PDF error from PDF replicas propagated to final observables.

We analyze these components separately and as a composition through the stages. When quoting a single band, we combine them, but we also display kinematics‑dependent residuals \(\Delta_{\rm acc}(x,k,Q)\) to make the precision component explicit.

With \(s(x,k; Q)\) in hand, we form observables by inserting Eq.~\eqref{eq:convolution_stage2} into Eq.~\eqref{eq:FUU_joint} and reconstruct the cross-sections in the same binning and with the same Jacobian \(E\,d\sigma/d^3q=(2\pi q_T)^{-1} d\sigma/(dq_T\,dy)\) used for the experimental presentations. The per‑\((x,Q)\) normalization ensures matching onto the collinear PDFs, and the small‑\(q_T\) working region avoids any explicit modeling of the \(Y\)-term tail.

\section{\label{sec:closure}Closure Tests}

We validate the momentum–space extraction with three closure tests that mirror the steps of the pipeline: a cross–section–level closure using pseudo–data generated with fixed–target kinematics; a closure on the convolution kernel \(S(q_T)\); and a closure on the intrinsic profile \(s(k)\). The latter two follow the inverse–integration workflow of Sec.~\ref{sec:methodology}. Once a DNN is trained, both \(S(q_T)\) and \(s(k)\) can be sampled on arbitrarily fine meshes in \(q_T\) or \(k\); they are therefore not limited by sparse experimental binning. An example of one of the basic cross–section closure tests is displayed in Fig.~\ref{fig:cross_section_closure} using experimental kinematics and errors, while the \(S(q_T)\) and \(s(k)\) closures appear in Fig.~\ref{fig:SqT_sk_closure}. We emphasize that the latter (Stage-II closure test) is a level-0 (noiseless) inverse-operator
validation designed to isolate inversion bias; statistical fluctuations are instead
propagated through the replica ensemble, which provides the higher-level uncertainty on
the reconstructed $s(x,k_\perp,Q)$ which is studied in the end-to-end closure tests.

\subsection{\label{sec4-1}cross-sections}

For the initial controlled test we adopt a simple generator with no explicit \(x\)– or \(Q\)–dependence for simplicity,
\begin{align}
  s_{\text{true}}(k) = \frac{1}{\pi}\,e^{-k^{2}},
\end{align}
which implies the analytic auto–convolution
\begin{align}
  S_{\text{true}}(q_T) = \frac{1}{2\pi}\,\exp\!\left(-\frac{q_T^{2}}{2}\right).
\end{align}
Pseudo–data are produced at E288–like kinematics (with the \(\Upsilon\) veto) using 
Eq. (\ref{eq:FUU_joint}) together with modern collinear inputs (e.g.\ NNPDF4.0). We then carry out the end‑to‑end fit, training directly on the triple–differential cross-sections and reconstructing the theory at bin centers with the invariant Jacobian of Eq.~\eqref{eq:inv}. Agreement between prediction and pseudo–data across \((q_T,Q_M,y)\) slices demonstrates a consistent forward map; representative comparisons are shown in Fig.~\ref{fig:cross_section_closure}.

\subsection{\label{sec4-2}\(S(q_T)\) closure (inverse integration)}

Given the trained model, we extract \(S(q_T;x_1,x_2,Q)\) by factoring out known kinematic prefactors and PDF combinations, and compare it against the truth computed from the generator via the same momentum–space operator used in the fit,
\begin{widetext}
\begin{align}
S(q_T;x_1,x_2,Q)=\int_{0}^{2\pi}\! d\phi \int_{0}^{k_{max}}\! dk\,k\;
s(x_1,k;Q)\,s\!\left(x_2,\sqrt{q_T^2+k^2-2q_Tk\cos\phi}\,;\,Q\right).
\label{eq:closure_S_integral}
\end{align}
\end{widetext}
The inference in this case is essentially continuous in $S(q_T;x_1,x_2,Q)$, limited only by computational resources.  We sample \(S(q_T)\) on a dense \(q_T\) grid (far denser than the experimental binning since we are in the general problem mapping a DNN model directly). The left panel of Fig.~\ref{fig:SqT_sk_closure} shows that the learned kernel reproduces the generator line shape within the quoted bands over the resolved range.

\subsection{\label{sec4-3}\(s(k)\) closure (inverse problem)}

We invert for the integrand by training the physics‑informed \(s(x,k,Q)\) against the \(S(q_T)\) labels from the previous step, enforcing the per‑\((x,Q)\) normalization \(\int d^2k\,s=1\) and smoothness priors. For the generator above the target is known analytically, so we test whether the trained \(s(k)\) recovers the original line shape:
\begin{align}
  s_{\text{true}}(k)=\frac{1}{\pi}\,e^{-k^2}
  \qquad \text{vs.} \qquad
  s_{\text{extr}}(k).
\end{align}
The right panel of Fig.~\ref{fig:SqT_sk_closure} demonstrates near‑perfect recovery across the resolved \(k\) window, with a mild systematic deviation at very small \(k\). That low‑\(k\) feature reflects the interplay of low‑\(q_T\) leverage, the \(k\)‑space taper controlling boundary effects, and finite angular sampling; it is included in the methodological error budget below.


\begin{figure}[ht!]
  \centering
  \includegraphics[width=1.0\columnwidth]{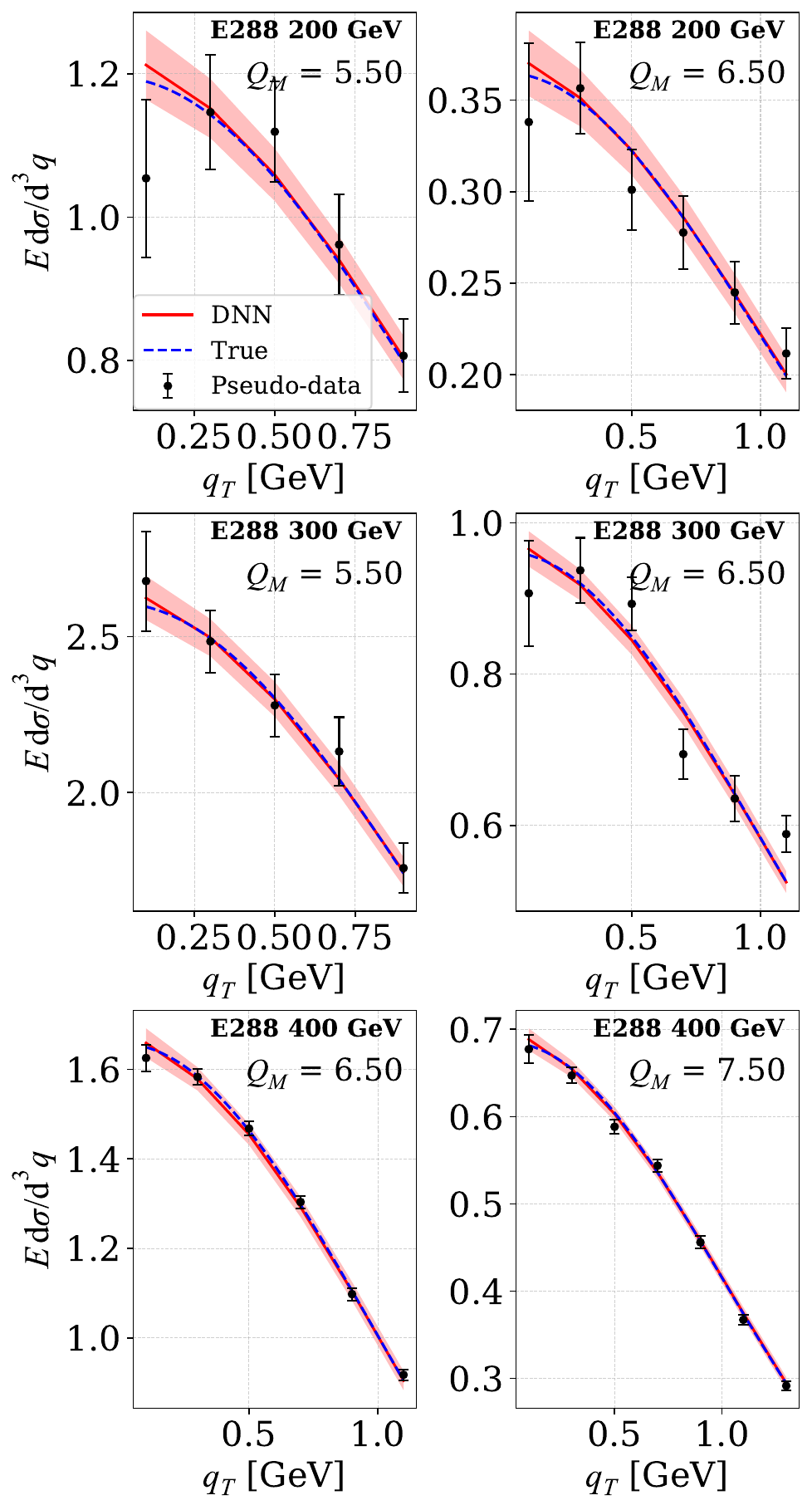}
  \caption{Cross‑section closure with pseudo‑data at E288‑like kinematics (in black) with the same scale of error from the experiment. The dashed line (in blue) show the pseudo‑data (true cross-sections); the solid line and the error band (in red) represent the `mean' and 1$\sigma$ DNN prediction (with 1000 replicas) respectively; reconstructed at bin centers using Eq.~\eqref{eq:inv}.}
  \label{fig:cross_section_closure}
\end{figure}

The residual $\Delta_{\rm acc}(x,k,Q)$ in $s(k)$ is a crucial diagnostic, as it quantifies the error associated with the inverse mapping. The aim is to achieve a near–perfect recovery of the input $s(k)$; any deviation reflects an uncertainty that must be incorporated into the total error budget. This component of the methodological uncertainty is intrinsic to inverse problems of this kind. 
The width of the algorithmic error distribution among replicas in the reconstructed 
\( s(k) \) should be negligible, since the mapping is direct and free from additional stochastic sources of error. 
The residual term \( \Delta_{\rm acc}(x,k,Q) \) must likewise be minimized by design, 
with any remaining quantifiable contribution propagated as a kinematically dependent bias 
in the fidelity of the reconstructed \( s(k) \).

\begin{figure*}[t]
  \centering
  \includegraphics[width=\textwidth]{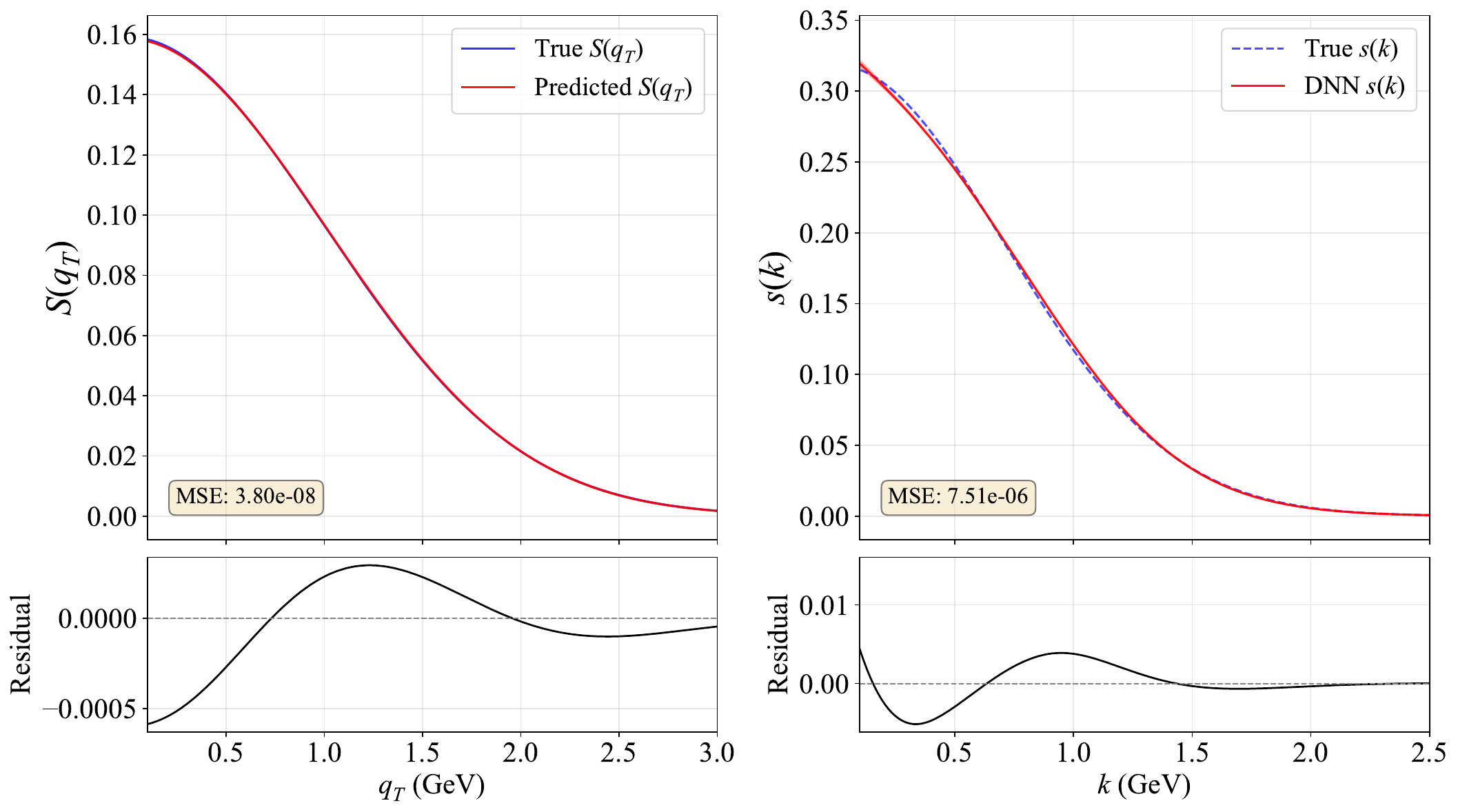}
  \caption{Closure for \(S(q_T)\) and \(s(k)\) using the inverse‑integration workflow (results with 1000 replicas). \textbf{Left:} \(S(q_T)\) comparison (learned vs.\ generator). \textbf{Right:} \(s(k)\) comparison (extracted vs.\ generator). Where shown, the shaded band indicates the \(1\sigma\) algorithmic spread across independent trainings.}
  \label{fig:SqT_sk_closure}
\end{figure*}

\subsection{\label{sec4-4}Recursive precision/accuracy accounting}

To fully quantify \emph{precision} from \emph{accuracy} and stabilize the inversion, we follow a recursive protocol: generate pseudo–data from a current generator \(G^{(r)}\); perform the cross‑section fit to obtain \(S^{(r)}\); reconstruct \(s^{(r)}\); compare with real data; update \(G^{(r)}\) and repeat. Iterations stop when pseudo–data match the real cross-sections (bin‑wise) within experimental errors across kinematics. The final, kinematics‑dependent \emph{accuracy} residual is given in Eq. (\ref{delta_acc}), and is reported alongside the \emph{precision} band obtained from the replica ensemble (shown as a \(1\sigma\) algorithmic band on \(S(q_T)\) and \(s(k)\)).

\subsection{\label{sec4-5}Methodological error}

We monitor four contributions in addition to the statistical replica spread. The \textit{algorithmic} component captures sensitivity to network initialization and optimizer schedule and appears as the \(1\sigma\) band around the mean in \(S(q_T)\) and \(s(k)\). The \textit{integration/inversion} component is obtained by switching evaluators (direct angular–radial quadrature versus auto‑convolution integration) and tightening/loosening \(N_\phi\), radial nodes, and \(k_{\max}\); the induced shift is quoted as a systematic. The \textit{experimental} component comes from cross‑section replicas propagated through the Stage‑I fit into the \(S\) labels, and the \textit{collinear‑PDF} component from PDF replicas used in the final observable reconstruction. To probe sensitivity to spectral content in \(k\), we also vary the generator around the baseline (e.g.\ width and tail deformations) and take the difference between the true and deduced means as a \textit{generator‑modeling} contribution. In the Gaussian test, this analysis yields a small low‑\(k\) deviation that remains subdominant to the quoted precision band.

\subsection{\label{sec4-6}Remarks on density and kinematic coverage}

Because \(S(q_T)\) and \(s(k)\) are produced from trained DNNs rather than directly from sparse detector binning, they can be evaluated on arbitrarily fine grids. We therefore present closure comparisons on dense \((q_T,k)\) meshes, while ensuring that anchors (normalization and low‑\(q_T\) curvature) are computed with the same quadrature used during training. This dense evaluation is crucial for diagnosing localized biases (e.g.\ the low‑\(k\) behavior noted above) and for defining kinematics‑aware accuracy residuals within the recursive protocol; see Figs.~\ref{fig:cross_section_closure} and \ref{fig:SqT_sk_closure}.

\section{\label{sec5:fits}Fits with Experimental Data}

The extraction of unpolarized TMDs from fixed-target DY data proceeds through a momentum–space pipeline that connects the measured cross-sections directly to the underlying transverse–momentum distributions using the same two stage and recursive methodology previously outline and demonstrated with the pseudo-data. We continue to work entirely in $k_\perp$ space, using a deep neural network (DNN) as a physics-informed \emph{integrand} for the transverse convolution. The procedure imposes per-$(x,Q)$ normalization and smoothness while minimizing functional bias by avoiding parametric Ans\"atze.

\subsection*{Step 1 — Cross-section replicas and learning $S(q_T;x_1,x_2,Q_M)$}
The starting point is the triple-differential cross-section data from the E288 and E605 experiments, binned in $(q_T,Q_M,y)$. We reconstruct $(x_1,x_2)$ at leading order in $q_T/Q_M$ and apply the standard cuts ($q_T<0.2 \; Q_M$ and the $\Upsilon$ veto: $9~\mathrm{GeV}<Q_M<11~\mathrm{GeV}$). For each dataset we generate ensembles of \emph{experimental replicas} by sampling the reported statistical uncertainties; full covariance matrices are not available and are therefore not used. A supervised model is then trained on the replicas to predict the structure function $F^{1}_{UU}$, from which we isolate the convolution kernel $S(q_T;x_1,x_2,Q_M)$ by dividing out known kinematic prefactors and the collinear PDF combinations (with PDF replicas used for uncertainty propagation). The loss is a weighted mean squared error that respects the experimental errors and kinematic coverage, and progressive training is employed for stability. The trained ensemble is projected onto a fine $(x_1,x_2,q_T,Q_M)$ grid to produce averaged $S$ labels with associated bands, which become the targets for the inversion in Step~2.

\subsection*{Step 2 — Reconstruction of $s(x,k_\perp;Q)$ in $k_\perp$ space}
To solve the inverse problem, we represent the intrinsic profile $s(x,k_\perp;Q)$ by a normalized DNN and embed it as the integrand of the transverse-momentum convolution, as described in Eq. \ref{eq:closure_S_integral}. The forward operator is implemented via a differentiable angular–radial quadrature inside the training loop, so the map from network parameters to observables remains smooth and trainable with gradient methods. We enforce the per-$(x,Q)$ normalization $\int d^2k\,s(x,k;Q)=1$ and mild curvature regularization, and we include light physics anchors (e.g., low-$q_T$ behavior and an integrated moment of $S$) to stabilize the inversion. The result is a nonparametric $s(x,k_\perp,Q)$ that reproduces the learned $S$ labels across the full kinematic range of the experimental data.

\subsection*{Step 3 - Recursion and Error Extraction}
The same recursive refinement previously described in the methodology is applied to stabilize the inversion and to separate precision from accuracy. Each outer iteration consists of: (i) fitting the cross-sections to obtain $S$ on a dense grid; (ii) reconstructing $s(x,k,Q)$ against those $S$ labels; (iii) generating pseudo-data at higher resolution from the current $s$ and refitting; and (iv) repeating until pseudo-data agree with the real cross-sections, bin-wise, within their reported errors. This loop reduces algorithmic variance, mitigates initialization sensitivity, and quantifies accuracy via kinematics-dependent residuals evaluated on matched pseudo-data. All fits are performed with the $\Upsilon$ veto and with the conventional $q_T<0.2\,Q_M$. No further variation is seen after two iterations within the experimental error. 

\subsection*{Summary of TMD assembly}
After convergence, the unpolarized TMDs are assembled as
\begin{align}
  f_{1,q/N}(x,k_\perp;Q,Q^2)\;=\; f_{1,q}(x;Q^2)\, s(x,k_\perp,Q),
  \label{eq:fit_summary}
\end{align}
with $f_{1,q}(x;Q^2)$ taken from modern global fits of the collinear PDF and $s(x,k_\perp ;Q)$ obtained from the inversion above. The ensemble of trained models propagates experimental and PDF uncertainties to $S$ and, subsequently, to $s$ and to the final cross-section predictions. Representative cross-section comparisons and TMD slices are shown in Figs.~\ref{fig:cs_result_288},\ref{fig:cs_result_605} and \ref{fig:tmds_x01_q45}.

\section{The DNN $S(q_T;x_1,x_2,Q_M)$ fit}
\label{sec:stage1-Sfit}

\noindent
In Stage~I we learn the momentum–space kernel $S(q_T;x_1,x_2,Q_M)$ directly from the
triple–differential DY cross-sections by factoring out the known kinematic prefactors and the
collinear PDF combinations, and training a compact DNN on the resulting target.\footnote{We
follow the small–$q_T$ setup and notation of Secs.~II–III and the Stage‑I summary there.}
The learned $S$ is then used in Stage~II to reconstruct the intrinsic profile $s(x,k_\perp,Q)$
via the differentiable inverse integration described next.

\subsection*{Inputs, targets, and prefactors}
For each experimental point we form two auxiliary PDF channels that collect the charge–weighted
$q\bar q$ and $\bar q q$ combinations at the bin scale $Q_M$,
\begin{align}
f_1(x_1,x_2) &\equiv
 \sum_{q=u,d,s} e_q^2\;  x_1x_2 \; f_{1,q}(x_1;Q_M^2)\,f_{\bar q}(x_2;Q_M^2),\\
f_1(x_2,x_1) &\equiv
 \sum_{q=u,d,s} e_q^2\; x_1x_2 \; f_{1,q}(x_2;Q_M^2)\,f_{\bar q}(x_1;Q_M^2),
\end{align}
evaluated with LHAPDF (using NNPDF4.0).  The narrow $Q_M$–bin integral entering the unit
conversion and Jacobian is computed analytically,
\begin{align}
&\int_{Q_M-\Delta/2}^{Q_M+\Delta/2}\!\frac{dQ'}{Q'^3}
 \\[2pt]
&= \tfrac12\!\left[(Q_M-\tfrac{\Delta}{2})^{-2}-(Q_M+\tfrac{\Delta}{2})^{-2}\right]. \nonumber
\end{align}
We denote the overall factor
\begin{align}
&K(Q_M)\nonumber
 \\[2pt]
&= \frac{(4\pi\alpha)^2}{18\pi}\,
   \Big[(Q_M-\tfrac{\Delta}{2})^{-2}
       -(Q_M+\tfrac{\Delta}{2})^{-2}\Big]
   (\hbar c)^2 .
\label{eq:KQM}
\end{align}
Given the measured cross-section $\left(\frac{d\sigma}{dq_T\, dQ_M\, dy}\right)$, the per–point \emph{regression target} removes
$K(Q_M)$ and leaves only $S$ times the PDF combinations,
\begin{align}
&\mathrm{S}_{\mathrm{calc}}(q_T;x_1,x_2,Q_M)
\equiv
   \frac{\left(\frac{d\sigma}{dq_T\, dQ_M\, dy}\right)}{K(Q_M) \big[f_1(x_1,x_2)
      +f_1(x_2,x_1)\big]}
\\[2pt]
&\approx
   S(q_T;x_1,x_2,Q_M)\,
   .\nonumber
\label{eq:SBtarget}
\end{align}
In practice we train on Monte~Carlo \emph{replicas} of $\left(\frac{d\sigma}{dq_T\, dQ_M\, dy}\right)^{\rm rep}$ drawn from the reported
uncertainties.\footnote{Full
covariance matrices are unavailable for E288/E605, so we propagate experimental errors via
replicas; PDF uncertainties are handled by sampling PDF replicas and repeating the fit.}

\subsection*{Network and forward map}
The model is a feed–forward network with inputs
$(q_T,x_1,x_2,Q_M)$ and a single output $\widehat S_{DNN}$ with a
\texttt{softplus} activation.  We then multiply $\widehat S_{DNN}$ by the two PDF channels and sum:
\begin{widetext}
\begin{equation}
\widehat{\mathcal{S}}(q_T;x_1,x_2,Q_M)
= \widehat S_{\rm DNN}(q_T;x_1,x_2,Q_M)\,
\big[f_1(x_1, x_2)
      +f_1(x_2, x_1)\big],
\label{eq:Sx}
\end{equation}
\end{widetext}
which establishes sum‑aggregation as the canonical permutation‑invariant operator \cite{Zaheer2017DeepSets}.
This implements, inside the network graph, the same PDF–weighted structure-function
decomposition used in the analytic presentation.  For proton–proton ($pp$) kinematics, the kernel is symmetric under the exchange $(x_1, x_2)\leftrightarrow(x_2, x_1)$.
In our implementation, this symmetry is enforced by evaluating both orderings through the explicit PDF channels in Eq.~\eqref{eq:Sx}.
While an explicit symmetrization penalty or input sorting can be applied, it is not required in the default fit.
Furthermore, imposing beam–target exchange symmetry, as done in our fit, introduces a potential bias when using non-hydrogen targets (as in E288 and E605) that is not quantified within our current estimate of systematic uncertainty and will be addressed in future studies.

The DNN architecture for $S$ is already described in Sec. \ref{subsec:3-A}. Typical training settings are: initial learning rate $2\times10^{-3}$,
batch size $64$, and up to $4000$ epochs run in two progressive stages (see below).

\subsection*{Loss, weights, and progressive training}
The objective is a weighted Mean Squared Error (MSE) over training rows $i$,
\begin{align}
\mathcal L_S \;=\; \frac{1}{N}\sum_{i=1}^{N} w_i\;
\big[\widehat{\mathrm{S}}_i - \mathrm{S}^{\rm rep}_i\big]^2,
\end{align}
with per–row weights $w_i$ used to emphasize under–represented panels in $(\text{dataset},Q_M)$.
Training proceeds progressively in $N_{\rm prog}=2$ stages, freezing earlier layers in the first
stage and unfreezing in the second.  We employ an early–stopping rule keyed to the crossover of
validation and training losses after a minimum burn‑in (here $100$ epochs) with patience and a
small tolerance on $(\mathrm{val}\!-\!\mathrm{train})$; a \texttt{ReduceLROnPlateau} callback gently
lowers the learning rate when the loss stalls.

\subsection*{From $\widehat{\mathrm{S}}$ to cross-sections and to the $S$ grid}
After training, predictions on any input batch are converted back to cross-sections via
\begin{align}
\widehat{\left(\frac{d\sigma}{dq_T\, dQ_M\, dy}\right)} \;=\; \widehat{\mathrm{S}}(q_T;x_1,x_2,Q_M)\;K(Q_M),
\end{align}
using the same $K(Q_M)$ as in Eq.~\eqref{eq:KQM}.  For the hand‑off to Stage~II we project the trained
model onto a fine grid in $(x_1,x_2,q_T,Q_M)$ and \emph{divide out the PDF channels} to recover
\begin{align}
\mathcal{S}(q_T;x_1,x_2,Q_M)
= \frac{\widehat{\mathcal{S}}(q_T;x_1,x_2,Q_M)}{f_1(x_1,x_2)+f_1(x_2,x_1)}\
\end{align}
thereby producing the $S$ labels and bands that drive the inverse $k_\perp$ fit in the next
section.  Uncertainties are propagated by repeating the Stage‑I training over experimental and
PDF replicas and taking the ensemble mean and dispersion on the grid.

\subsection*{Position in the pipeline}
This Stage‑I fit isolates the kernel $S(q_T;x_1,x_2,Q_M)$ from data while respecting the same
kinematic prefactors and PDF structure used in the analytic formalism, staying natively in
momentum space.  The resulting $S$ surfaces, with their replica bands, become the targets for
Stage~II’s differentiable inverse convolution, where we reconstruct the normalized intrinsic
profile $s(x,k_\perp;Q)$ that reproduces $S$ across the full kinematic phase space learned by the DNN.

\section{Differentiable $k_\perp$ inverse Convolution}
\label{sec:inverse-kperp-film}

Before detailing the forward model used in the inversion, we note that
reconstructing a transverse profile from its projected $q_T$ convolution
is an ill-conditioned inverse problem.
During the initial prototyping stage we therefore introduce a small set
of auxiliary priors—soft pins, moment anchors, curvature and tail
stabilizers—to condition the optimization and to facilitate stable
architecture and hyperparameter selection through recursive closure
tests. These auxiliary terms are used only to guide model configuration
and are subsequently removed, except for the mass normalization which is
retained by definition to enforce PDF–TMD consistency along with the cross section components of the loss.
The final reported solutions are therefore obtained without removable
regularization terms; the temporary priors serve only to stabilize early
training and do not determine the physical extraction.
\subsection{Forward model in $k_\perp$ for inversion}
In the small–$q_T$ DY regime the unpolarized structure function is written as 
Eq. (\ref{eq:FUU_joint}) with the beams exchanged for $pp$ kinematics as appropriate.  Our inverse problem is:
given triple–differential cross-sections in $(q_T,Q_M,y)$ (after the standard cuts),
reconstruct a \emph{single} nonparametric profile $s(x,k;Q)$ such that its
auto–convolution reproduces the learned $S$ and, through
$F^{1}_{UU}$, the measured cross-sections.

Two structural constraints are imposed throughout the fit: (i) per–point PDF
matching, implemented via the normalization
\begin{align}
\int d^2\!k\, s(x,k,Q)
&= 2\pi \int_{0}^{\infty}\! dk\,k\, s(x,k,Q)
 = 1,\\& \forall(x,Q) \nonumber.
\end{align}
and (ii) monotone falloff in $k$, enforced by the parameterization below.

\subsection{Monotone integrand via a FiLM–conditioned $\lambda$–network}
\label{sec:lambda-net}

Rather than regressing $s$ directly, we learn its \emph{logarithmic slope}
$\lambda(x,k,Q)\!\ge\!0$ with a compact stack of \emph{feature‑wise linear modulation} (FiLM)‑gated residual blocks~\cite{He2016ResNet} and define
\begin{align}
s(x,k,Q)&=\exp\!\Big[-\!\int_{0}^{k}\lambda(x,\kappa,Q)\,d\kappa\Big],
\label{eq:monotone-s}
\end{align}

\begin{align}
s(x,0,Q)&=1.
\label{eq:monotone-s-1}
\end{align}
With $\lambda\!\ge\!0$, $s$ is positive and monotonically decreasing in $k$ by construction; the
exponential‑of‑integral map is evaluated \emph{inside} the computation graph and is standard in
differentiable‑programming settings~\cite{Chen2018NeuralODE,Mildenhall2020NeRF}. This parameterization
enforces qualitative shape constraints without hand‑crafted ans\"atze~\cite{Amos2017ICNN,You2017DeepLattice,Durkan2019NSF}.

The network input is $(x,k,Q)$. To ease the learning of the radial dependence we expand the scalar
$k$ into a fixed feature vector
\[
\psi(k)\equiv \big(k,\;k^2,\;\sqrt{k+\varepsilon},\;\log(1{+}k)\big),
\]
with a small $\varepsilon>0$ to avoid the square‑root singularity at $k=0$. The initial hidden state is
$h^{(0)}=\phi\!\big(W^{(0)}\psi(k)\big)$, where $\phi=\tanh$ and $W^{(0)}\!\in\!\mathbb{R}^{W\times 4}$.

Non‑separable $(x,Q)$ dependence is injected through FiLM
\cite{Perez2018FiLM,Dumoulin2017Conditional}. Let $c(x,Q)$ denote a small conditioning stream; for each
block $\ell=1,\dots,L$ it produces a pair of modulation vectors
\begin{equation}
(\gamma_\ell(x,Q),\,\beta_\ell(x,Q)) \;=\; g_\ell\!\big(c(x,Q)\big)\in \mathbb{R}^{W}\times\mathbb{R}^{W}.
\label{eq:film-gate}
\end{equation}
Given the incoming hidden state $h^{(\ell)}\!\in\!\mathbb{R}^{W}$, the FiLM‑gated \emph{residual} block
(with width $W$) computes
\begin{align}
u^{(\ell)} &= \phi\!\big(W^{(\ell)}_{1}\,h^{(\ell)}\big) \in \mathbb{R}^{W},
\label{eq:film-u}\\
\tilde u^{(\ell)} &= \gamma_\ell(x,Q)\;\odot\;u^{(\ell)} \;+\; \beta_\ell(x,Q) \in \mathbb{R}^{W},
\label{eq:film-affine}\\
h^{(\ell+1)} &= \phi\!\Big(W^{(\ell)}_{2}\,\tilde u^{(\ell)} \;+\; h^{(\ell)}\Big) \in \mathbb{R}^{W}.
\label{eq:film-res}
\end{align}
Here $W^{(\ell)}_{1},W^{(\ell)}_{2}\!\in\!\mathbb{R}^{W\times W}$ are learnable weight matrices; $\odot$ denotes
the Hadamard (elementwise) product; and the additive term $+\,h^{(\ell)}$ is the residual (skip) connection
\cite{He2016ResNet}. For complete clarity, Eq. \eqref{eq:film-u}–\eqref{eq:film-res} can be written componentwise as
\begin{align}
u^{(\ell)}_j &= \tanh\!\Big(\sum_{i=1}^{W}(W^{(\ell)}_{1})_{ji}\,h^{(\ell)}_i\Big),~j=1,\dots,W,\\
\tilde u^{(\ell)}_j &= \gamma_{\ell,j}(x,Q)\;u^{(\ell)}_j \;+\; \beta_{\ell,j}(x,Q),\\
h^{(\ell+1)}_j &= \tanh\!\Big(\sum_{i=1}^{W}(W^{(\ell)}_{2})_{ji}\,\tilde u^{(\ell)}_i \;+\; h^{(\ell)}_j\Big).
\end{align}
In our default implementation we also consider a scalar‑gate variant where
$\gamma_\ell,\beta_\ell\in\mathbb{R}$ are \emph{scalars} per block (and per sample), broadcast across features;
Eq.~\eqref{eq:film-affine} then reduces to $\tilde u^{(\ell)}=\gamma_\ell u^{(\ell)}+\beta_\ell \mathbf{1}$ with
$\mathbf{1}\in\mathbb{R}^{W}$ the all‑ones vector. We use $\gamma_\ell=\mathrm{softplus}(\cdot)$ and
$\beta_\ell=\tanh(\cdot)$ to ensure a positive gain and bounded shift.

After $L$ blocks, the slope is produced by a one‑neuron head with a \texttt{softplus} activation,
\begin{align}
\lambda(x,k,Q) \;&=\; \mathrm{softplus}\!\big(w_{\rm out}^{\!\top} h^{(L)} + b_{\rm out}\big),
\\ w_{\rm out}\!&\in\!\mathbb{R}^{W},\;\; b_{\rm out}\!\in\!\mathbb{R},
\label{eq:lambda-head}
\end{align}
which guarantees $\lambda\!\ge\!0$ and, through Eq.~\eqref{eq:monotone-s}, 
implies $s(x,k;Q)>0$ and monotone
decrease in $k$. The per‑$(x,Q)$ mass constraint $\int d^{2}k\,s(x,k;Q)=1$ is enforced in the physics
layer that performs the $k_\perp$ quadrature (Sec.~\ref{sec:diff-evaluator}).

The FiLM gate Eq. \eqref{eq:film-affine} implements a kinematics‑dependent affine transformation of the radial
features, enabling $(x,Q)$‑dependent \emph{deformations} of the $k$ profile without factorizing variables or
inflating parameters. The residual skip in Eq. \eqref{eq:film-res} provides an identity gradient pathway, which
stabilizes deep stacks of blocks. The map $\lambda\mapsto s$ is differentiable, so gradients from both the
pointwise $S$ fidelity term and the cross‑section–level $\chi^2$ propagate through the FiLM gates and the
quadrature layer end‑to‑end.

The tuned FiLM-conditioned residual DNN architecture is the following:  Among the three inputs $(x, k, Q_M)$, the primary variable $k$ 
undergoes feature expansion into four representations $(k,\, k^2,\, \sqrt{k},\, 
\log(1+k))$, which are projected into a 64-dimensional hidden state via a 
$\tanh$-activated dense layer. In parallel, a conditioning subnetwork processes 
$(x, Q_M)$ through two ReLU-activated layers of width 32 to produce per-block 
FiLM parameters---a $\mathrm{softplus}$-activated scale $\gamma$ and a 
$\tanh$-activated shift $\beta$---one pair per residual block. The trunk consists 
of 4 residual blocks, each comprising two dense layers of width 64 with $\tanh$ 
activations; between them, the FiLM affine transformation $\gamma \odot 
\mathbf{h} + \beta$ modulates the intermediate representation, allowing $x$ and 
$Q_M$ to dynamically gate the network's response to $k$. Residual skip 
connections stabilise training across depth. The final layer is a single 
$\mathrm{softplus}$ unit with a calibrated bias initialisation, guaranteeing a 
positive output $\lambda(x, k, Q_M) \geq 0$ throughout the domain. The maximum network comprises 35,081 trainable parameters.


\subsection{Differentiable $k_\perp$ evaluator and normalization}
\label{sec:diff-evaluator}
We embed Eq.~\eqref{eq:S_def_joint} and Eq. ~\eqref{eq:monotone-s} into a custom
\emph{differentiable quadrature layer} that is part of the computational graph \cite{Baydin2018AD}:
\begin{itemize}
\item \textbf{Adaptive radial grid:} We place $R_{\rm nodes}{\simeq}4096$ nodes
$k_i = k_{\max}\tanh(\beta\,\xi_i)/\tanh\beta$ with $\xi_i$ uniform in $[0,1]$ and
$\beta\sim3$, concentrating resolution at small~$k$.
\item \textbf{Azimuthal integration:} The $\varphi$ integral uses $N_\varphi$ uniform points
($N_\varphi=64$–$128$; see ablations), and for each $(q_T,k_i,\varphi_j)$ we evaluate
$k'=\sqrt{q_T^2+k_i^2-2q_T k_i\cos\varphi_j}$ and linearly interpolate $s(x_2,k',Q)$ on the
$\{k_i\}$ grid.
\item \textbf{Trapezoid in $k$ and tapering:} The $k$–integral uses a trapezoid rule on the
adaptive grid. To suppress endpoint artifacts we multiply the \emph{left} profile by a
cosine taper $t(k)$ that turns on smoothly for $k \gtrsim 0.85\,k_{\max}$ and track a
``tail–mass'' beyond $0.9\,k_{\max}$.
\item \textbf{Normalization:} For every $(x,Q)$ we evaluate the mass
$m_0[\,s\,]=2\pi\!\int dk\,k\, s$ on a high–order Gauss–Legendre grid and \emph{normalize}
$s\mapsto s/m_0[s]$ (``left'' normalization). A symmetric option applies the same to the
right profile (``both''); in practice we use ``left'' to reduce redundancy while retaining
stability.
\end{itemize}
Because Eq.~\eqref{eq:monotone-s} is an ordered cumulative integral, the discrete Jacobian is
lower triangular: on the grid one has
$\partial s(k_j)/\partial \lambda(k_i) = -\,s(k_j)\,\Delta k_i$ for $i\le j$ and $0$ otherwise.
Backpropagation through the $\varphi$–integral and the $k$–trapezoid therefore yields stable,
banded gradients all the way to the FiLM parameters.


Let $m_2[f]\equiv 2\pi\!\int dk\,k^3 f(k)$ denote the second radial moment:
For a convolution of two \emph{independent, normalized, isotropic} profiles one has
$m_2[S]=m_2[s_1]+m_2[s_2]$; for identical beams this gives $m_2[s]=\tfrac12 m_2[S]$.
We therefore penalize $(m_2[s]-\tfrac12 m_2[S])^2$ with $m_2[S]$ estimated from the
learned $S$ via $2\pi\!\int dq\,q^3 S(q)$.  Two low–$q_T$ anchors stabilize the origin:
$S(0)$ and a pin on $S''(0)$ obtained by local quadratic fits at the first $q_T$ nodes.
We also include a curvature prior in $k$ and a tail–mass penalty beyond the taper.\\

\subsection{Direct cross–section supervision and the inverse loss}
\label{sec:cs-loss}
In addition to the pointwise $S$ loss, the model is supervised \emph{directly} by the
experimental triple–differential cross-sections.  We define an aggregator $G_i$ that maps
each training row $i$ (with inputs $X_i=(x_{1,i},x_{2,i},q_{T,i},Q_{M,i})$) to the
corresponding experimental kinematic bin $j=(\text{dataset},Q_{M},q_T)$ and forms the cross-section,\footnote{For brevity we denote the measured (binned) cross-section as $A_j^{\exp}$ and the prediction as $A_j^{\text{pred}}$.}
\begin{widetext}
\begin{equation}
A^{\rm pred}_j \;=\; \sum_{i\in\mathcal{B}_j} G_i\;
S_\theta(q_{T,i};x_{1,i},x_{2,i},Q_{M,i}),
\label{eq:asym-agg}
\end{equation}

\begin{equation}
G_i = \Bigg[\sum_{q=u,d,s} e_q^2\Big(
f_{q}(x_{1,i};Q_M^2)f_{\bar q}(x_{2,i};Q_M^2)
+ f_{\bar q}(x_{1,i};Q_M^2)f_{q}(x_{2,i};Q_M^2)\Big)\Bigg]
\;\mathcal{K}(Q_M),
\label{eq:cs-agg}
\end{equation}
\end{widetext}
where the factor $\mathcal{K}(Q_M)$ collects the tree–level hard prefactor and is define in Eq.~(\ref{eq:KQM}). 
 With measurements $(A_j^{\rm exp}\!\pm dA_j)$ we define the cross-section loss
\begin{equation}
\mathcal{L}_{\rm CS} \;=\; \frac{1}{J}\sum_{j=1}^{J} w_j\;
\left(\frac{A^{\rm pred}_j - A^{\rm exp}_j}{dA_j}\right)^2,
\label{eq:Lcs}
\end{equation}
where $w_j$ are optional per–panel weights (e.g.\ to focus selected $(\text{dataset},Q_M)$
panels). The pointwise $S$ loss on training rows $i\in\mathcal{T}$,
$\mathcal{L}_{S}=\frac{1}{|\mathcal{T}|}\sum_{i\in\mathcal{T}} \omega_i\,[S_\theta(X_i)-S^{\rm targ}(X_i)]^2$,
uses inverse–variance weights $\omega_i\propto 1/\sigma_i^2$. In the nominal optimization we set $\omega_i=1$.

The total objective combines the data-consistency terms with the physics-based regularizers described above:
\begin{widetext}
\begin{align}
\mathcal{L}_{\text{total}} &=
   \mathcal{L}_{S}
 + \lambda_{\text{CS}}\mathcal{L}_{\text{CS}}
 + \lambda_{\text{mass}}\,[m_0[s]-1]^2
 + \lambda_{m_2}\,[m_2[s]-\tfrac12 m_2[S]]^2 \nonumber\\[2pt]
&\quad
 + \lambda_{0}\,[S(0)-S_0]^2
 + \lambda_{2}\,[S''(0)-S_2]^2
 + \lambda_{\text{curv}}\,\|\partial_k^2 s\|_{w(k)}^2
 + \lambda_{\text{tail}}\,m_0[s;\,k>k_{\text{tail}}] .
\label{eq:Ltotal}
\end{align}
\end{widetext}
with $\|\cdot\|_{w(k)}$ a curvature norm reweighted by a learned, nonnegative profile $w(k)$
(see below), and $k_{\rm tail}\simeq 0.9\,k_{\max}$.  All terms are computed inside the same
differentiable layer, so gradients from $\mathcal{L}_{\rm CS}$ flow through the
aggregator Eq.~\eqref{eq:cs-agg} and the $k_\perp$ evaluator to the FiLM parameters.

We employ a two–phase schedule matched to the implementation. 
\emph{Phase~I (pretrain):} minimize the pointwise $S$ data–fidelity loss 
$\mathcal{L}_{S}$ (weighted MSE; optionally log–MSE) to stabilize the forward map. 
\emph{Phase~II (fine–tune):} switch to full‑batch Adam on the combined objective
$\mathcal{L}_{\text{total}}$ in Eq.~\eqref{eq:Ltotal} with a nonzero cross‑section weight 
$\lambda_{\text{CS}}\!\in\![0.1,0.3]$, for $\mathcal{O}(10^2$–$10^3)$ steps. 
Unless stated otherwise we use \(\lambda_{\text{CS}}=0.20\). 
All symbols $\lambda$ in~\eqref{eq:Ltotal} are \emph{hyperparameters} that control the
relative strength of physics anchors and regularizers with respect to the data–fidelity terms.  We treat the anchor weights as fixed across replicas and datasets to avoid hyper‑overfitting; only model parameters are optimized.
For clarity we list them below together with the quantities they weight:\vspace{-2mm}
\begin{itemize}\setlength\itemsep{2pt}
  \item \textbf{$\lambda_{\text{CS}}$} multiplies the cross-section level $\chi^2$ term $\mathcal{L}_{\text{CS}}$, which compares the
        \emph{differentiably aggregated} prediction to the binned experimental cross-sections.
  \item \textbf{$\lambda_{\text{mass}}$} penalizes deviations from per–$(x,Q)$ unit mass:
        \(m_0[s]\equiv 2\pi\!\int_0^\infty\! dk\,k\,s(x,k,Q)\) and the penalty is \([m_0[s]-1]^2\).
  \item \textbf{$\lambda_{m_2}$} anchors the second moment of $s$ to the small–$q_T$ information in $S$:
        \(m_2[s]\equiv 2\pi\!\int_0^\infty\! dk\,k^3 s(x,k,Q)\), matched to \(\tfrac12 m_2[S]\) extracted from the learned
        $S(q_T)$ (the factor $\tfrac12$ is the Jacobian relating the two moments).
  \item \textbf{$\lambda_{0}$} and \textbf{$\lambda_{2}$} softly pin the $q_T\!\to\!0$ behavior of $S$ via 
        \([S(0)-S_0]^2\) and \([S''(0)-S_2]^2\), where $(S_0,S_2)$ are low‑$q_T$ anchors inferred from the $S$ fit.
        (These anchors may be set to zero in ablations.)
  \item \textbf{$\lambda_{\text{curv}}$} is a Tikhonov‑type smoothness weight on the radial profile
        \(\|\partial_k^2 s\|_{w(k)}^2\), with
        \begin{align}
        \|\partial_k^2 s\|_{w(k)}^2 \;\equiv\; \int_{0}^{k_{\max}}\! dk\; w(k)\,[\partial_k^2 s(x,k,Q)]^2,
        \end{align}
        where \(w(k)\!\ge\!0\) is a user‑specified (or learned) nonnegative weight and \(k_{\max}\) is the
        maximal $k$ covered by the evaluator.
  \item \textbf{$\lambda_{\text{tail}}$} penalizes excess probability in the far tail,
        \(m_0[s;\,k>k_{\text{tail}}]\equiv 2\pi\!\int_{k_{\text{tail}}}^{\infty}\! dk\,k\,s(x,k,Q)\), to confine mass
        within the region constrained by data.
\end{itemize}
In our default configuration we found stable performance with \(\lambda_{m_2}\!\sim\!5\times10^{-2}\),
\(\lambda_{\text{curv}}\!\sim\!2\times10^{-4}\), \(\lambda_{\text{tail}}\!\sim\!10^{-3}\), and very small or
vanishing $(\lambda_{0},\lambda_{2})$; the unit‑mass constraint is also enforced by the evaluator’s
normalization, so \(\lambda_{\text{mass}}\) can be kept small.\footnote{Exact scalings can differ with grid
resolution $k_{\max}$ and the choice of $w(k)$; we keep all $\lambda$’s fixed across replicas and datasets.}

To make the smoothness term focus where the data are most informative, we optionally adapt
\(w(k)\) from the gradient of the $S$ fidelity term:
\begin{align}
w_{\text{new}}(k)
&\;\propto\;
\frac{\big|\partial\mathcal{L}_{S}/\partial s(x,k,Q)\big|^{\alpha}}
     {\big(|\partial_k^{2}s(x,k,Q)|^{\beta}+\epsilon\big)},
\label{eq:PARupdate}\\[3pt]
&\text{clipped to } [1,\,w_{\max}],
\nonumber
\end{align}
with $(\alpha,\beta,w_{\max})=(0.9,1.0,6.0)$ by default and $\epsilon\!>\!0$ for stability; $w_{\text{new}}$ is
renormalized to unit mean. In the cross-section driven setup used for the main results.


\subsection{Discretization and numerical settings}
\label{sec:numerics}
Unless otherwise noted we use $k_{\max}=5~\mathrm{GeV}$, $R_{\rm nodes}=4096$,
$N_\varphi=64$–$128$, a taper onset at $0.85\,k_{\max}$ with a tail mask beyond
$0.9\,k_{\max}$, and $512$–point Gauss–Legendre quadrature for the normalization
and anchor moments.  These settings were verified to be numerically converged
for the resolved $(q_T,k)$ windows and are consistent with the ablations reported
elsewhere in the paper.

\subsection{Target anchors}
\label{sec:m2-derivation}

The moment anchors in the loss function connect the intrinsic transverse–momentum
profile $s(x,k;Q)$ to the experimentally accessible structure kernel $S(q_T;x_1,x_2,Q)$ through
their statistical moments.  To make this connection explicit, let
$\mathbf{K}_{1,2}$ denote the two independent, isotropic transverse–momentum vectors
carried by the partons in the incoming hadrons, each distributed according to the
normalized radial density $s_{1,2}(k)$ with $k\!=\!|\mathbf{K}|$.  The observable
transverse momentum of the lepton pair is their vector sum,
\begin{equation}
\mathbf{Q} \;\equiv\; \mathbf{K}_1 + \mathbf{K}_2,
\end{equation}
so that $S(q_T)$ is the radial probability density of $\mathbf{Q}$ with
$q_T\!=\!|\mathbf{Q}|$.

Because $\mathbf{K}_1$ and $\mathbf{K}_2$ are statistically independent and isotropic,
their mean–squared magnitudes add,
\begin{equation}
\langle Q^2\rangle
  = \langle K_1^2\rangle + \langle K_2^2\rangle.
\end{equation}
In two–dimensional polar coordinates one may write the second radial moment of any
normalized isotropic density $f(k)$ as
\begin{align}
m_2[f] \;\equiv\; 2\pi \!\int_0^\infty\! dk\, k^3 f(k)
                \;=\; \langle K^2\rangle_f .
\end{align}

Applying this to the convolution relation between $s_{1,2}$ and $S$ gives
\begin{equation}
m_2[S] \;=\; m_2[s_1] + m_2[s_2].
\end{equation}
For identical beams ($s_1\!=\!s_2\!=\!s$) this reduces to
\begin{equation}
m_2[s] \;=\; \tfrac{1}{2}\,m_2[S],
\label{eq:m2relation}
\end{equation}
which is exactly the relation used as the second-moment
anchor in~\eqref{eq:Ltotal}.

\paragraph{Role in the training schema.}
Equation~\eqref{eq:m2relation} supplies a physically interpretable constraint that ties
the learned intrinsic width of $s(x,k;Q)$ to the measured width of the observable
$S(q_T;x_1,x_2,Q)$.  In practice, $m_2[S]$ is computed from the
Stage-I kernel prediction via
\begin{align}
m_2[S] = 2\pi\!\int_0^\infty dq_T\, q_T^3 S(q_T),
\end{align}
and the penalty term
$[m_2[s]-\tfrac{1}{2}m_2[S]]^2$ in Eq.~\eqref{eq:Ltotal} stabilizes the inversion by
preventing $s(x,k;Q)$ from drifting to unphysical widths that would otherwise yield
degenerate $S(q_T)$ spectra.
Together with the low-$q_T$ anchors on $S(0)$ and $S''(0)$,
this constraint fixes the overall scale and curvature of the reconstructed distribution,
ensuring that the learned $s(x,k;Q)$ reproduces both the integrated and differential
behavior of the measured cross-sections while remaining consistent with the
auto-convolution relation in Eq. \ref{eq:closure_S_integral}.

\subsection{End–to–end differentiability}
\label{sec:end-to-end}

The entire model is trained as a single differentiable computation that links the learnable
parameters of the network to the final observable cross-sections. Denoting by
$\theta$ the collective set of \emph{trainable parameters} of all neural components
(weights and biases of the FiLM-conditioned $\lambda$-network, the curvature weight
profile $w(k)$ if enabled, and the output head), the forward map can be written schematically as


\begin{widetext}
\begin{align}
\label{eq:forward-map}
\theta 
&\;\mapsto\;
\lambda_\theta(x,k,Q)
\;\mapsto\;
s_\theta(x,k,Q) \xrightarrow{\text{quadrature layer}}
S_\theta(q_T;x_1,x_2,Q)
\xrightarrow{\text{PDFs \& binning}}
A^{\text{pred}}_j(\theta)
\end{align}
\end{widetext}

Each arrow corresponds to a differentiable operator inside the computational graph:
\begin{itemize}\setlength\itemsep{3pt}
  \item[$\bullet$] 
  $\lambda_\theta(x,k,Q)$ is produced by the FiLM-gated residual network described in
  Sec.~\ref{sec:lambda-net}; it depends explicitly on $(x,k,Q)$ and implicitly on the learnable
  parameters $\theta$.
  \item[$\bullet$]
  $s_\theta(x,k,Q)$ is obtained through the cumulative integral
  $s=\exp[-\!\int_0^{k}\lambda\,d\kappa]$, implemented differentiably on the adaptive
  $k$-grid within the same graph. The Jacobian
  $\partial s(k_j)/\partial \lambda(k_i)$ is lower-triangular, ensuring stable gradient
  flow during backpropagation.
  \item[$\bullet$]
  The \emph{quadrature layer} performs the radial and azimuthal integrations that map
  $(s_1,s_2)\!\to\!S(q_T)$ via the convolution; its discretized form retains exact differentiation with
  respect to $\theta$.
  \item[$\bullet$]
  The final stage combines $S_\theta$ with charge-weighted parton densities
  $f_q(x_i;Q^2)$ and experimental bin integrals to yield
  $A^{\text{pred}}_j(\theta)$, the theoretical cross-section in bin $j$
  as defined in Eq. ~\eqref{eq:asym-agg}.  
\end{itemize}

Because every stage is differentiable, gradients of the total loss
$\mathcal{L}_{\text{total}}(\theta)$ propagate seamlessly from the experimental
$\chi^2$ and pointwise $S$ terms down to the fundamental model parameters:
\[
\frac{\partial \mathcal{L}_{\text{total}}}{\partial \theta}
=
\frac{\partial \mathcal{L}_{\text{total}}}{\partial A^{\text{pred}}}
\frac{\partial A^{\text{pred}}}{\partial S_\theta}
\frac{\partial S_\theta}{\partial s_\theta}
\frac{\partial s_\theta}{\partial \lambda_\theta}
\frac{\partial \lambda_\theta}{\partial \theta}.
\]
This chain rule—executed automatically by the deep-learning framework—ensures that both
pointwise $S$ supervision and the cross-section–level $\chi^2$ drive the FiLM-conditioned
$\lambda$-network toward an $s(x,k;Q)$ whose auto-convolution reproduces the measured
$S(q_T;x_1,x_2,Q)$ while simultaneously satisfying the physics anchors and normalization
constraints.

Conceptually, Eq.~\ref{eq:forward-map} realizes an \emph{end-to-end differentiable
inverse problem} in momentum space: the model learns to invert the $k_\perp$ convolution
operator directly, without factorized \textit{Ansätze} or a detour through the
$b_T$ representation~\cite{Bracewell1999}.  The differentiable structure allows
the optimizer to exploit analytic gradients instead of finite-difference updates,
yielding stable and physically interpretable solutions. Such end-to-end differentiable
inversion schemes have become canonical in modern scientific machine learning for
ill-posed inverse problems~\cite{Arridge2019Inverse,Adler2018PrimalDual}.

As a robustness check, we performed systematic ablation studies to assess the role of all
auxiliary constraints introduced for early-stage stabilization. The essential cross section component of the loss and mass normalization
constraint are retained by definition, since it enforces PDF–TMD consistency; however,
all other removable terms—including soft pins, moment anchors, curvature priors, and
tail penalties—are used only during the initial prototyping phase to identify stable
architectures and hyperparameters through recursive closure validation.

Once convergence and stability are established, these auxiliary penalties are removed
one by one in a controlled sequence of closure tests, and the final model is obtained
without them. Across the resolved $(x,k_\perp,Q)$ domain, eliminating these removable
constraints leads to changes below $\sim 3\%$ in both magnitude and shape of the
extracted TMDs, with no statistically significant impact on the cross-section
description. This confirms that the final reported solution is determined by the
data and the inverse-convolution constraint itself, while the auxiliary terms serve
only as numerical stabilizers during model conditioning and do not materially bias
the extraction.

\begin{figure*}[ht!]
    \centering
    \includegraphics[width=2.05\columnwidth]{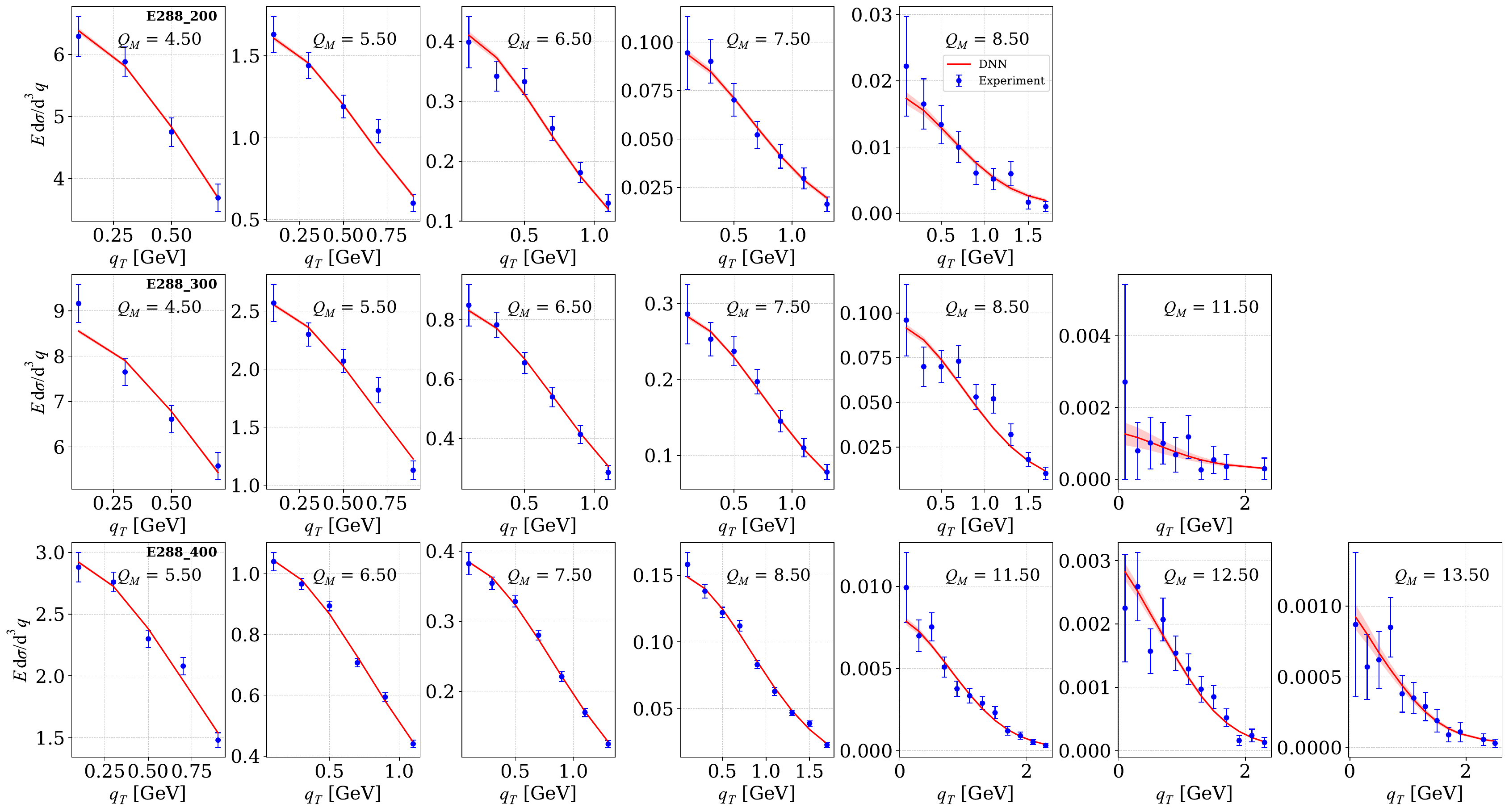}
    \caption{Representative cross-section comparison after the joint $s(x,k_\perp,Q)$ extraction. Points with error bars are E288 data (with standard cuts); the curve and band are the DNN prediction and its $1\sigma$ uncertainty obtained by propagating the replica ensemble through Eq.~\eqref{eq:fit_summary}, as such only the propagated experimental errors are shown.}
    \label{fig:cs_result_288}
\end{figure*}

\begin{figure*}[ht!]
    \centering
    \includegraphics[width=1.5\columnwidth]{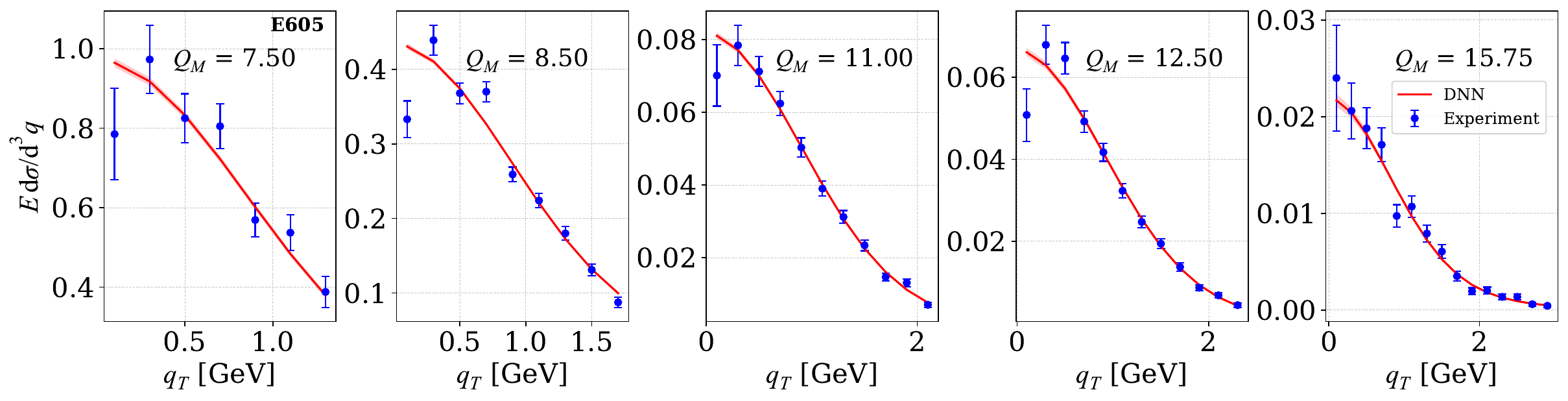}
    \caption{Representative cross-section comparison after the joint $s(x,k_\perp,Q)$ extraction. Points with error bars are E605 data (with standard cuts); the curve and band are the DNN prediction and its $1\sigma$ uncertainty obtained by propagating the replica ensemble through Eq.~\eqref{eq:fit_summary}, as such only the propagated experimental errors are shown.}
    \label{fig:cs_result_605}
\end{figure*}

\begin{figure*}[ht!]
    \centering
    \includegraphics[width=0.92\textwidth]{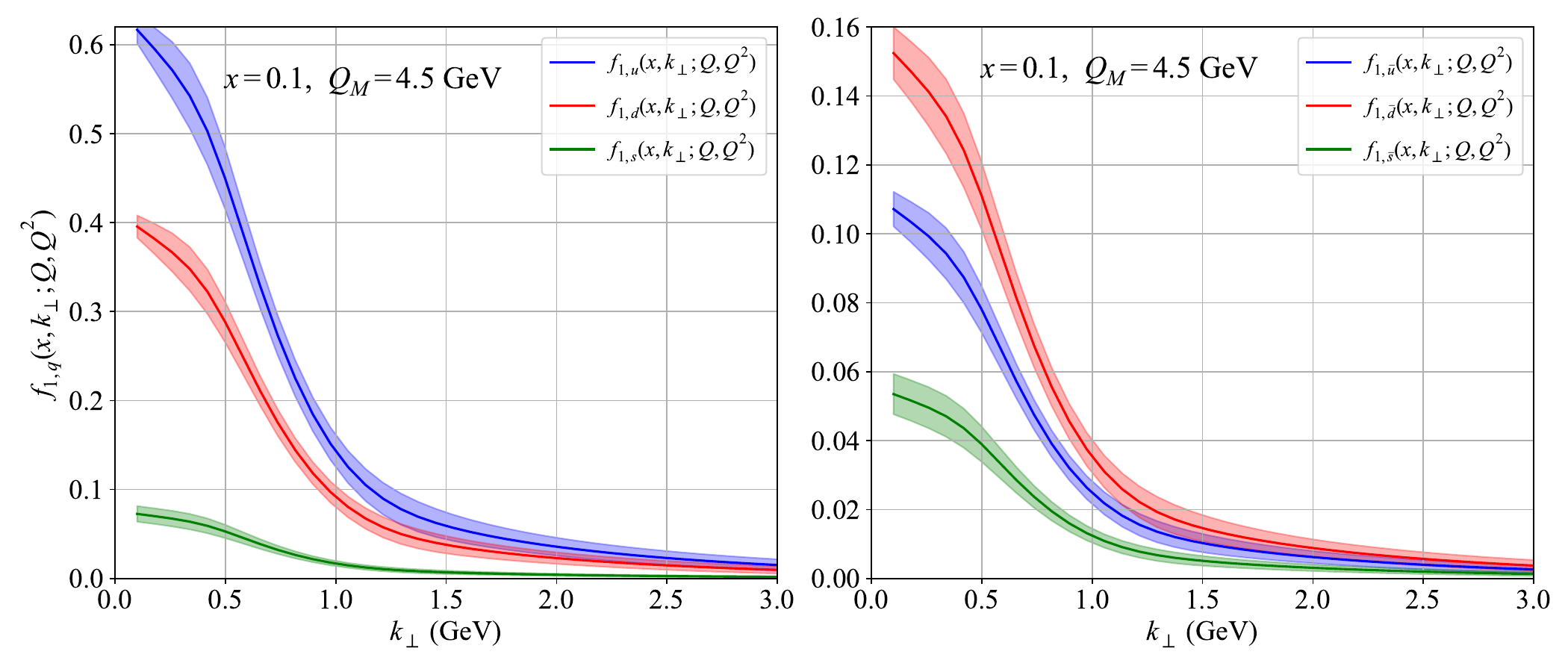}
    \caption{Unpolarized TMDs reconstructed at $x=0.1$ and $Q=4.5~\mathrm{GeV}$ using Eq.~\eqref{eq:fit_summary}. Curves show the central values for selected flavors; shaded bands indicate the propagated $1\sigma$ uncertainties from experimental replicas (Step~1) and collinear PDF replicas. All error contributions are shown.}
    \label{fig:tmds_x01_q45}
\end{figure*}

\section{Results}\label{sec:results}
With the momentum–space pipeline of Secs.~\ref{sec5:fits}–\ref{sec:inverse-kperp-film} and the joint representation
\(
f_{1,q/N}(x,k_\perp;Q,Q^2)=f_{1,q}(x;Q^2)\,s(x,k_\perp;Q)
\),
we extract the unpolarized TMDs for all quark-flavors in $SU(3)_{\text{flavor}}$ $(u,d,s,\bar{u},\bar{d},\bar{s})$ across the E288/E605 phase space.
All results shown apply the standard kinematic cuts (including the \(\Upsilon\) veto) and $q_T < 0.2 ~ Q$,
where the small‑\(q_T\) restriction that emphasizes the TMD regime. The
end‑to‑end fit reproduces the measured cross-sections panel‑by‑panel and then inverts for
the intrinsic profile \(s(x,k_\perp;Q)\) via the differentiable \(k_\perp\) convolution described
in Sec.~VII. Representative cross‑section comparisons appear in Fig.~\ref{fig:cs_result_288} (for E288) and Fig.~\ref{fig:cs_result_605} (for E605).

\subsection*{Flavor dependence at a fixed scale}
\begin{figure*}[ht!]
  \centering
  \includegraphics[width=1.0\textwidth]{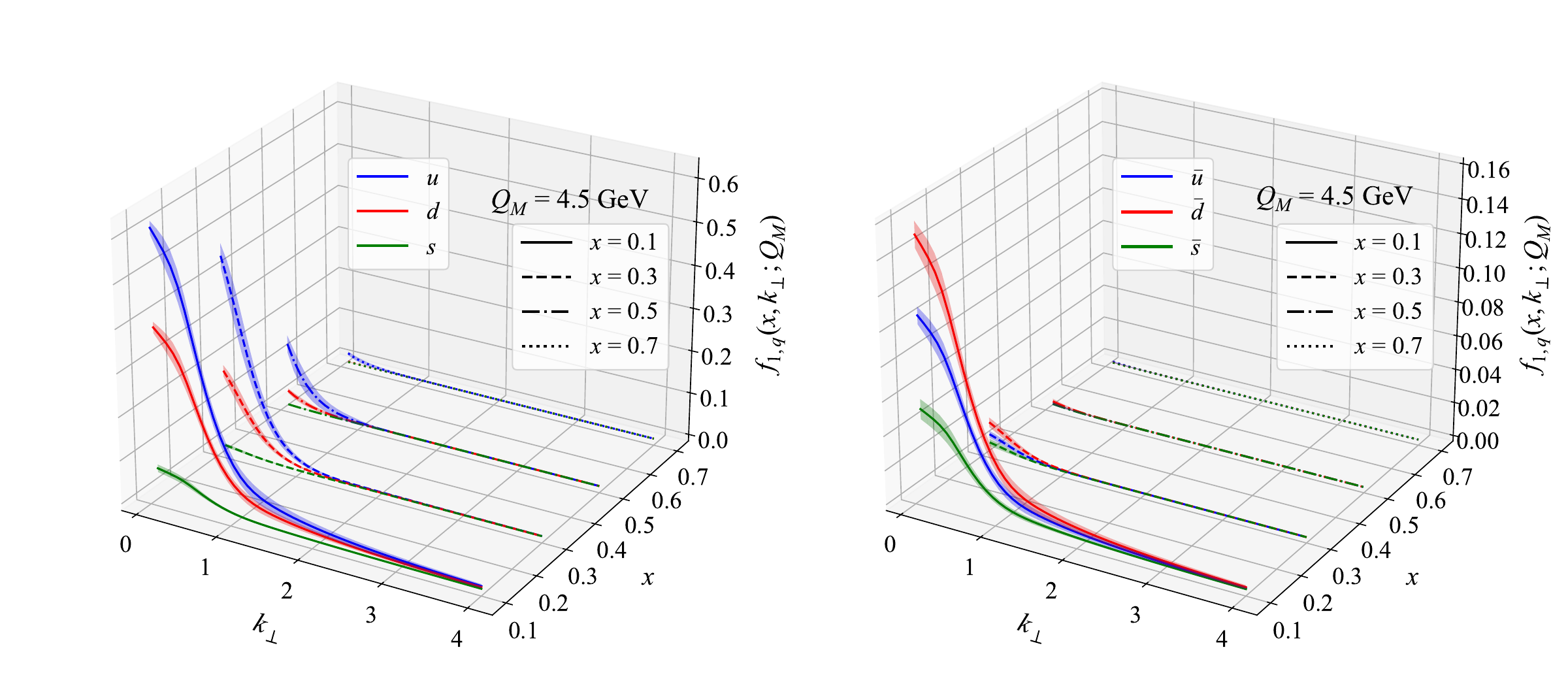}
\caption{
Transverse-momentum–dependent parton distributions
\( f_{1,q}(x,k_\perp;Q_M) \) at \( Q_M = 4.5~\mathrm{GeV} \),
shown for \(u\), \(d\), and \(s\) quarks (left) and antiquarks (right)
as functions of Bjorken-\(x\) and transverse momentum \(k_\perp\).
The distributions are reconstructed from the joint profile
\( s(x,k_\perp;Q) \) and assembled using Eq.~\eqref{eq:fit_summary}. 
Line styles correspond to fixed-\(x\) slices, illustrating the evolution
of the transverse-momentum width with \(x\).  All error contributions are shown.
}
  \label{fig:u3D_Q45}
\end{figure*}

\begin{figure*}[ht!]
  \centering
\includegraphics[width=1.0\textwidth]{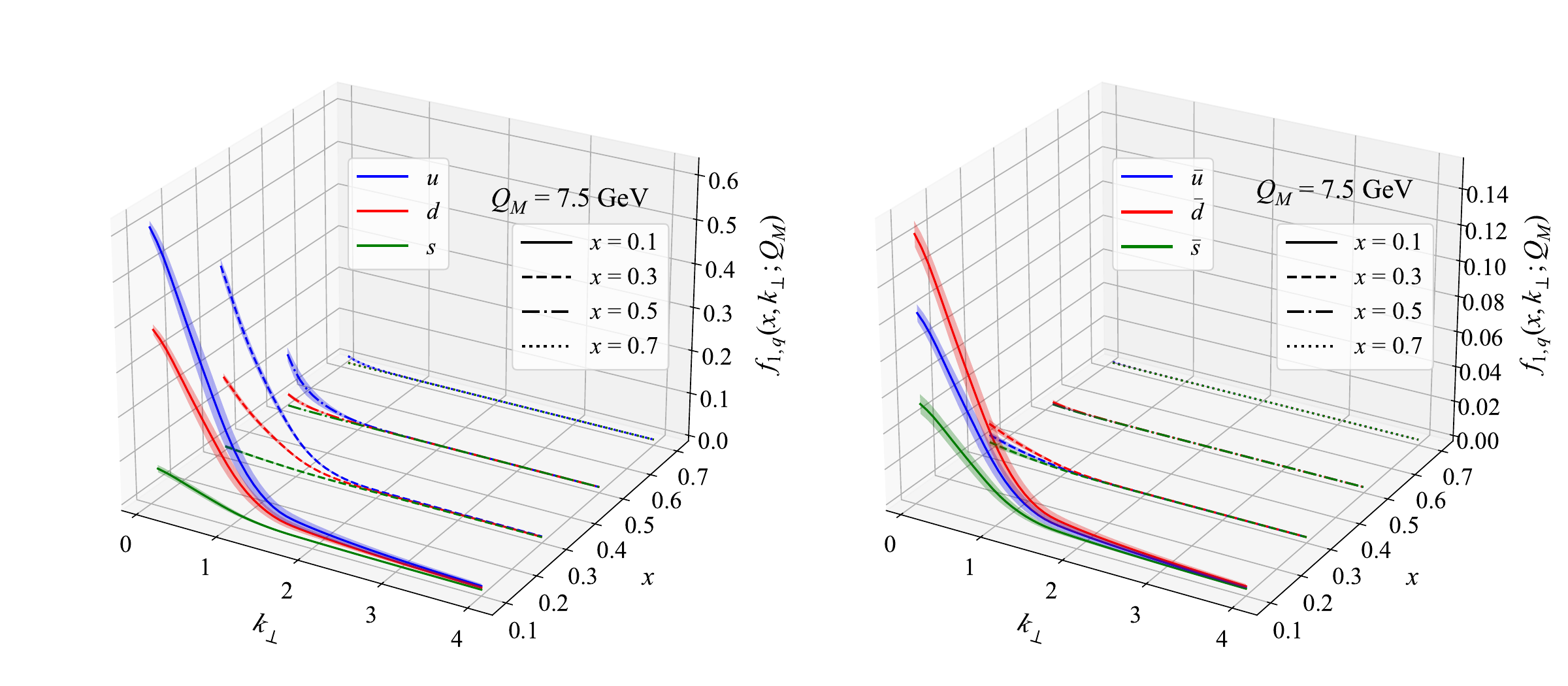}
\caption{
Transverse-momentum–dependent parton distributions
\( f_{1,q}(x,k_\perp;Q_M) \) at \( Q_M = 7.5~\mathrm{GeV} \),
shown for \(u\), \(d\), and \(s\) quarks (left) and antiquarks (right)
as functions of Bjorken-\(x\) and transverse momentum \(k_\perp\).
The distributions show mild 
\(k_\perp\) broadening over the fixed-target lever arm $Q_M$.
  All error contributions are shown.}
  \label{fig:u3D_Q75}
\end{figure*}

At fixed \((x,Q)\) the reconstructed TMDs exhibit the expected flavor hierarchy without
imposing a factorized Ansatz. Figure \ref{fig:tmds_x01_q45} shows \(f_{1,q}(x,k_\perp;Q,Q^2)\) (left) at \(x=0.1\) and
\(Q=4.5~\text{GeV}\): the \(u\) distribution is largest, followed by \(d\) and then \(s\), with all
curves falling smoothly in \(k_\perp\). The normalization \(\int d^2\!{\bf k_\perp}\,s(x,k_\perp;Q)=1\)
enforces PDF matching in Eq.~\ref{eq:pdf-matching}; the band reflects the replica ensemble propagated
through the two-stage fit. The $f_{\bar q}(x,k_\perp;Q,Q^2)$ distribution
for antiquarks (right panel) shows $f_{\bar d} > f_{\bar u}$. To visualize the global behavior across \((x,k_\perp)\) at a fixed scale, we show
the \(u\)‑quark surfaces at \(Q=4.5~\text{GeV}\) and \(Q=7.5~\text{GeV}\) in Figs.~\ref{fig:u3D_Q45} and~\ref{fig:u3D_Q75},
respectively for quarks (left) and anti-quark (right) in each case. These plots make explicit the joint \(x\)–\(k_\perp\) structure learned
from the data in momentum space.

The extraction captures a mild but systematic $Q$ dependence directly in $k_\perp$ space.
Because the fixed-target Drell--Yan data that we used in this paper (E288, E605) probe only a limited lever arm in the hard
scale, the resulting evolution of the intrinsic transverse-momentum profile is expected to be
modest in direct $(x,k_\perp)$ slices.
As the scale increases the TMDs exhibit a redistribution of strength toward larger $k_\perp$
and a corresponding suppression of the small-$k_\perp$ peak, while the per-$(x,Q)$ normalization
of $s$ preserves the overall mass of the profile.
This trend can be discerned by comparing the TMD plots at $Q=4.5$ and $7.5~\text{GeV}$
(Figs.~\ref{fig:u3D_Q45} and~\ref{fig:u3D_Q75}), and is quantified more clearly by the
moment-based width $\sigma_k(x,Q^2)$ shown in Fig.~\ref{fig:u-rms}.
The observed broadening is consistent with expectations from transverse-momentum resummation
in the small-$q_T$ domain, but here it is inferred empirically from the data within the
momentum-space inversion, rather than imposed through a predefined $b_T$-space evolution kernel.
Unlike $b_T$-space formulations where the Sudakov exponent makes the perturbative/nonperturbative
transition explicit, our present implementation does not attempt such a separation; instead it
learns an effective $Q$ dependence from data within the fixed-target lever arm by design.  The framework
can be coupled to a standard Sudakov/matching prior when advantageous.

\begin{figure}[ht!]
  \centering
\includegraphics[width=0.5\textwidth]{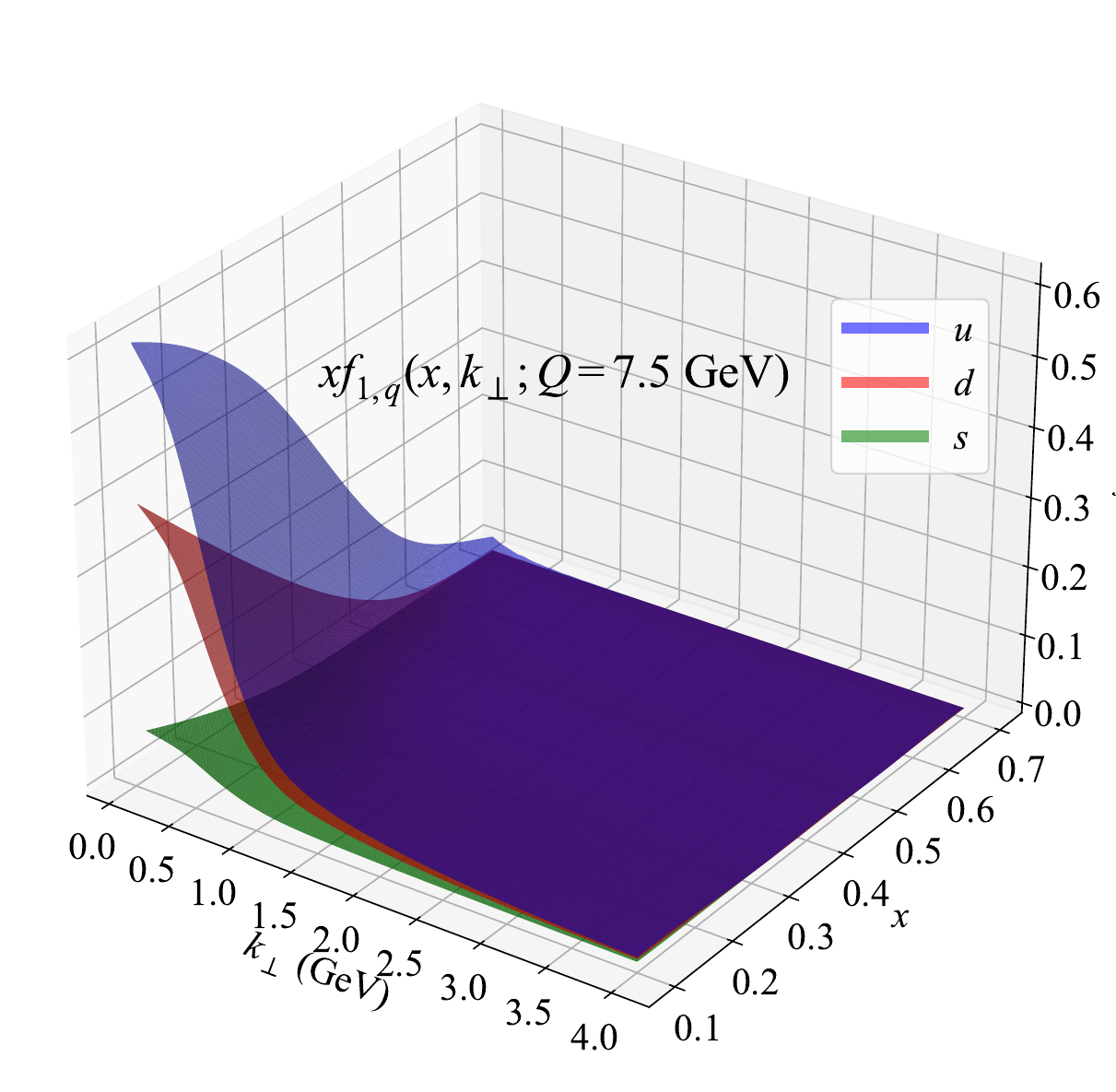}
\caption{Reconstructed $(x,k_\perp)$ surfaces of the unpolarized TMDs $f_{1,q}(x,k_\perp;Q,Q^2)$ for
$q=u,d,s$ at fixed $Q=7.5~\mathrm{GeV}$, shown to illustrate the joint $(x,k_\perp)$ structure
and flavor hierarchy at a single scale.
}
  \label{fig:u-surface}
\end{figure}

Figure~\ref{fig:u-surface} displays the reconstructed
$u$-quark transverse-momentum--dependent parton distribution
$f_{1,u}(x,k_\perp;Q)$ at a fixed scale $Q=7.5~\mathrm{GeV}$ across the
$(x,k_\perp)$ plane.
The surface is obtained directly in momentum space from the
joint profile $s(x,k_\perp;Q)$, assembled according to
Eq.~\ref{eq:fit_summary}, and represents the replica-averaged mean of the full
Monte-Carlo ensemble propagated through the differentiable convolution.
At this fixed scale, the distribution exhibits a smooth, monotonic falloff in
$k_\perp$, together with a gradual suppression at large~$x$ driven by the
collinear-PDF normalization.
The purpose of this figure is to illustrate the non-factorized $(x,k_\perp)$
structure learned from the data at a single scale, rather than to demonstrate
scale evolution.
The $Q$ dependence of the extracted transverse structure is assessed by comparing
fixed-$Q$ reconstructions at different scales and is quantified more directly by
the RMS width $\sigma_k(x,Q^2)$ shown in Fig.~\ref{fig:u-rms}.
The surface also demonstrates that the neural-network--based mapping preserves
positivity and per-$(x,Q)$ normalization while retaining the data-driven shape of
the extracted TMD.
The color map encodes the magnitude of $f_{1,u}(x,k_\perp;Q)$, with lighter tones
corresponding to larger probability density at small~$k_\perp$.

\begin{figure}[ht!]
  \centering
\includegraphics[width=0.5\textwidth]{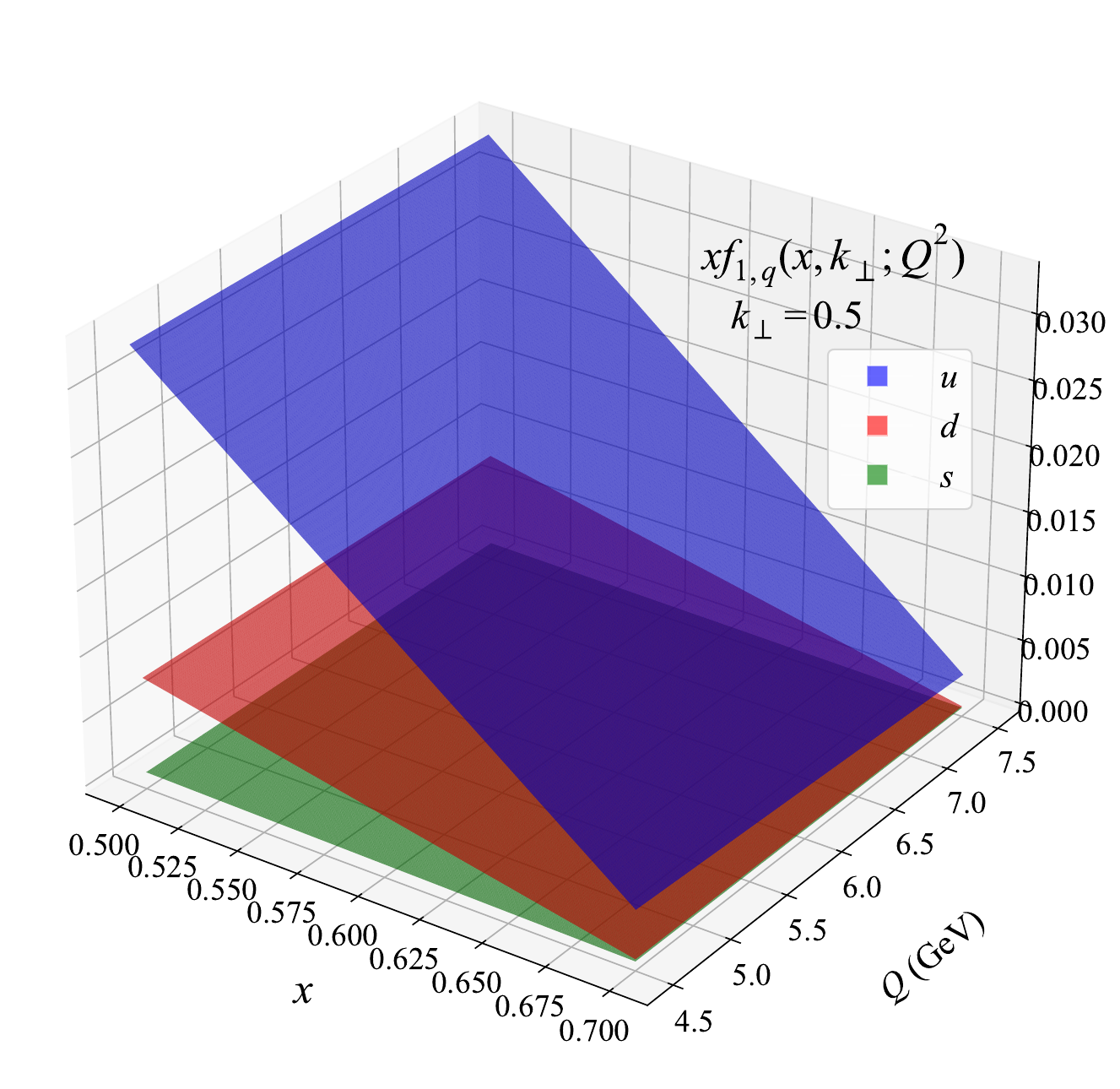}
\caption{
Reconstructed $(x,Q)$ surfaces of the unpolarized TMD at fixed representative
values of $k_\perp$. The figure illustrates the joint dependence on $Q$ and $x$ in
the kinematic region where the fixed-target data provide the strongest constraints.
}
  \label{fig:uds-evol}
\end{figure}

\begin{figure}[ht!]
  \centering
\includegraphics[width=0.5\textwidth]{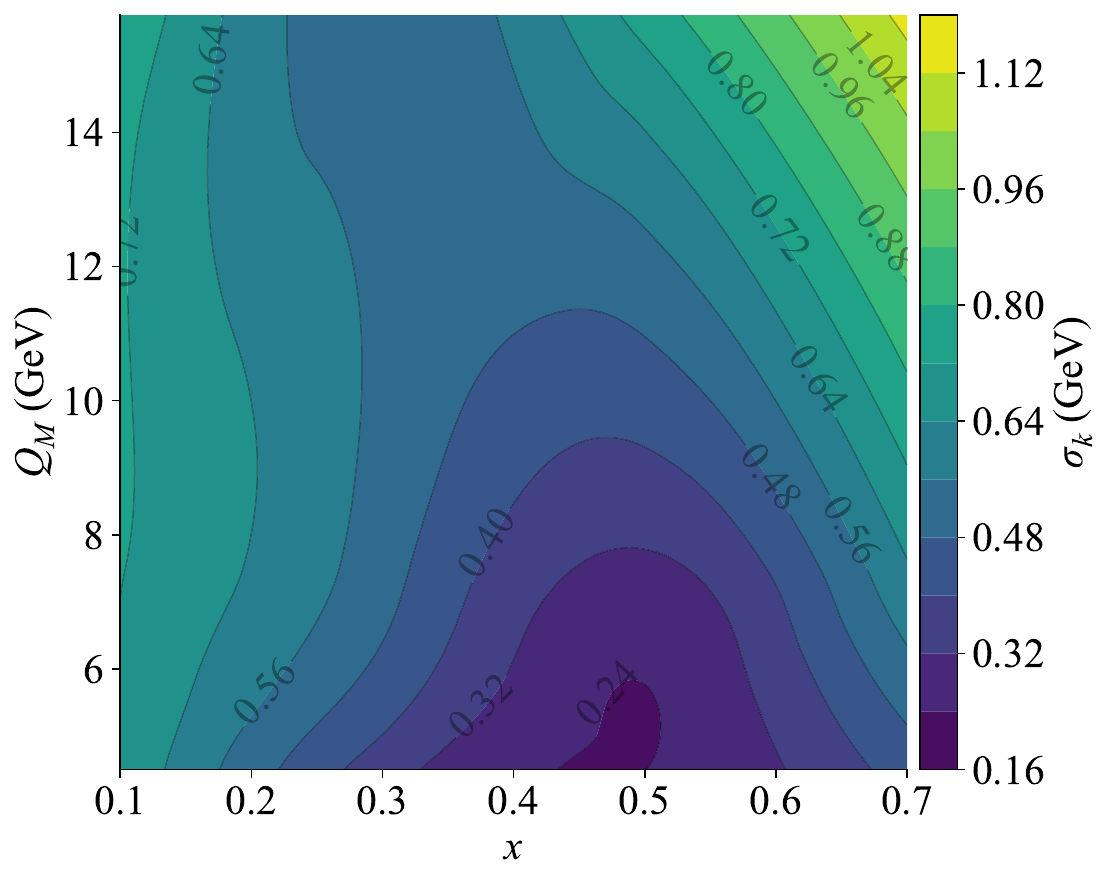}
\caption{
Root-mean-square (RMS) transverse-momentum width $\sigma_k(x,Q_M)$ of the extracted
unpolarized $u$-quark TMD, computed from replica-averaged normalized moments as
$\sigma_k=\sqrt{\langle k_\perp^2\rangle-\langle k_\perp\rangle^2}$.
The field is shown over the full $(x,Q_M)$ domain covered by the E288 and E605 kinematics
used in the analysis; so values displayed in that band reflect the smooth interpolation
of the learned model.
Across the fixed-target lever arm, $\sigma_k$ exhibits a dominant dependence on $x$ and a
modest but systematic increase with $Q_M$ that is most apparent at large~$x$, consistent with
the limited range in hard scale available in these data and with the fact that no explicit
Sudakov kernel is imposed.
}

  \label{fig:u-rms}
\end{figure}

A complementary view of the evolution is provided at fixed transverse momentum:
Fig.~\ref{fig:uds-evol} shows \(u,d,s\) surfaces versus \((x,Q)\) at \(k_\perp\approx0.5~\text{GeV}\). We focus on moderate-to-large values of $x$, where the E288/E605 data have the greatest
constraining power; lower-$x$ regions are better addressed in global analyses that
include collider data.
The pattern across flavors tracks the interplay between Sudakov broadening and the
collinear baseline \(f_{1,q}(x;Q^2)\), with the net change at fixed \(k_\perp\) varying with \(x\)
(as anticipated from DGLAP‑driven \(x\) evolution of the collinear inputs) while remaining smooth by construction of \(s(x,k_\perp;Q)\).

To compress the $(x,k_\perp)$ information into a single, scale–sensitive observable
we form the root–mean–square (RMS) transverse‑momentum width

\begin{align}
\sigma_k(x,Q) &\equiv
   \sqrt{\langle k_\perp^2\rangle - \langle k_\perp\rangle^2}\,
\end{align}
where $\langle k_\perp^2\rangle$ and $\langle k_\perp\rangle$ are defined as,
\begin{align}
    \langle k_\perp^2\rangle &\equiv
   \frac{\displaystyle \int d^2\!{\bf k_\perp}\,(k_\perp^2 )\,w(x,k_\perp;Q)}
        {\displaystyle \int d^2\!{\bf k_\perp}\,w(x,k_\perp;Q)}\,,
\end{align}
and,
\begin{align}
    \langle k_\perp\rangle &\equiv
   \frac{\displaystyle \int d^2\!{\bf k_\perp}\,(k_\perp )\,w(x,k_\perp;Q)}
        {\displaystyle \int d^2\!{\bf k_\perp}\,w(x,k_\perp;Q)}\,.
\end{align}

Here we use as weight either the fully assembled unpolarized TMD
$f_u(x,k_\perp;Q,Q^2)$ or, equivalently, the normalized profile
$s(x,k_\perp;Q)$, since $\int d^2\!{\bf k_\perp}\,s(x,k_\perp;Q)=1$ by construction.%
\;This follows from the joint momentum–space representation
$f_{1,q/N}(x,k_\perp;Q,Q^2)=f_{1,q}(x;Q^2)\,s(x,k_\perp;Q)$
and the per–$(x,Q)$ normalization imposed in the inverse‑integration stage, see Eqs.~\ref{eq:perpoint_norm}–\ref{eq:S_def_joint} and the assembly step summarized in Eq.~\ref{eq:cs-setup}.

Figure~\ref{fig:u-rms} shows $\sigma_k(x,Q_M)$ for the $u$ flavor.
Over the fixed-target kinematic window probed by E288 and E605,
$\sigma_k(x,Q_M)$ exhibits a pronounced dependence on $x$ and only a mild,
systematic variation with $Q_M$ mostly visible at large $x$, consistent with the limited hard-scale
lever arm of these data.
The dominant $x$ dependence reflects nontrivial correlations between
$x$ and the intrinsic transverse-momentum profile learned by the
momentum-space inversion, while the weak $Q_M$ dependence encodes a modest
redistribution of transverse momentum across scales within the fitted
small-$q_T$ domain.
Because $\sigma_k$ is constructed from normalized moments of the
reconstructed distribution, it is insensitive to the overall collinear
normalization and isolates genuine shape information.
In this sense, the $\sigma_k(x,Q_M)$ field provides a compact scalar
mapping of the transverse width learned directly from data, rather
than a stand-alone determination of perturbative Sudakov evolution.

\subsection*{Uncertainty characterization}
\label{subsec:uncertainties}

As detailed in Secs.~\ref{sec3-3} and~\ref{sec4-4}, uncertainties are propagated through a
Monte-Carlo replica workflow augmented by a dedicated closure-based estimate of
method-induced effects in the inverse step.  Here we \emph{summarize} the adopted
error components and the way they enter the final bands, and we report a compact
phase-space--averaged budget in Table~\ref{tab:uncertainty_budget}.  Terminology
and metrics follow the generalized uncertainty framework of
Ref.~\cite{Keller:UncProp:2025}.

We organize the total uncertainty into four contributions:
\emph{(i) experimental} (cross-section uncertainties),
\emph{(ii) collinear inputs} (PDF uncertainty entering $f_{1,q}(x;Q^2)$),
\emph{(iii) algorithmic/numerical} (training stochasticity and evaluator settings),
and \emph{(iv) methodological} (bias and spread associated with the inverse
reconstruction of the normalized profile $s(x,k_\perp;Q)$ and with other modeling
choices not captured by the replica loop).

\paragraph{Experimental replicas:}
Because full covariance matrices are not available, experimental uncertainties are propagated by generating Monte Carlo
replicas of each reported cross-section point using the quoted per-bin (stat.\,+ uncorrelated syst.) uncertainties.
In addition, where an overall (luminosity/normalization) uncertainty is quoted for a dataset, we treat it as a fully
correlated systematic by introducing a replica-dependent nuisance normalization factor $N_d^{(k)}$ common to all bins
$i$ in dataset $d$:
\begin{align}
A^{(k)}_{i,d} &= N_d^{(k)}\,A^{\rm data}_{i,d} \;+\; r^{(k)}_{i,d}\,\sigma^{\rm ptp}_{i,d}, \\
N_d^{(k)} &= 1 + r^{(k)}_{d}\,\delta^{\rm norm}_{d},
\qquad r^{(k)}_{d},\,r^{(k)}_{i,d}\sim \mathcal{N}(0,1).
\label{eq:mc_replicas_norm}
\end{align}
Here $\sigma^{\rm ptp}_{i,d}$ denotes the point-to-point (uncorrelated) uncertainty used for bin $i$ and
$\delta^{\rm norm}_{d}$ is the quoted relative normalization uncertainty for dataset $d$.
For E605 we use $\delta^{\rm norm}=15\%$ and include the quoted $10\%$ point-to-point systematic in quadrature with the
published bin uncertainties \cite{Moreno1991}, while for E288 we use $\delta^{\rm norm}=25\%$ \cite{Ito1981} as a correlated normalization component
(common within each E288 subset used in the fit).%

\paragraph{Collinear (PDF) inputs:}
Collinear-PDF uncertainty is propagated by sampling a PDF replica in sync with
each experimental replica and assembling
$f_{1,q/N}(x,k_\perp;Q,Q^2)=f_{1,q}(x;Q^2)\,s(x,k_\perp;Q)$ for that sample.

\paragraph{Algorithmic/numerical:}
We quantify residual stochastic and numerical scatter by repeating otherwise
identical trainings with different random seeds and by coherently varying key
evaluator settings (e.g.\ quadrature orders and grid parameters).  This component
is assessed \emph{without} experimental smearing and is kept subdominant by
construction.

\paragraph{Methodological error:}
To quantify method--induced systematics independently of experimental and PDF
uncertainties, we evaluate a two-part budget: a \emph{methodological accuracy}
(bias under noiseless closure) and a \emph{methodological precision} (spread under
controlled variations of modeling choices).
Let $\Omega$ denote a kinematic grid in $(x,k,Q)$ with $k\equiv|\mathbf{k}_\perp|$ and
empirical sampling weights $\rho(x,Q)$ (proportional to the training density).
Define the pointwise residual and an $L^2$ accuracy metric
\begin{align}
&\Delta_{\rm acc}(x,k,Q) \equiv s_{\rm extr}(x,k,Q)-s_{\rm true}(x,k,Q),\\
&\mathcal{A}[s] \equiv \nonumber
\Bigg[\sum_{(x,Q)\in\Omega_{xQ}}\rho(x,Q)\;\\
&\times2\pi\!\!\sum_{k\in\Omega_k}k\,\Delta_{\rm acc}(x,k,Q)^2\,\Delta k \Bigg]^{\!1/2},
\end{align}
which discretizes $\int\!dx\,dQ\,\rho(x,Q)\int\!dk\,(2\pi k)\,\Delta^2$ on the evaluator grid.

\emph{Stage~I (kernel $S$):} we scan the $S$--network architecture and training
hyperparameters (depth/width, activations, optimizer,
learning rate/batch size, early stopping) to minimize the validation
$\mathcal{L}_{S}$ while monitoring generalization (train vs.\ val). The scan is
embedded in a closure loop (noiseless pseudo-data) until $\mathcal{A}[s]$ and
the binwise $\chi^2$ stabilize across admissible settings. Residual biases at
this point are recorded as Stage-I methodological \emph{accuracy}; the dispersion
across the admissible hyperparameter set defines the Stage-I methodological
\emph{precision}.

\emph{Stage~II (inverse integrand $s$):} we perform a stricter recursive
calibration of the differentiable evaluator and anchor weights:
\begin{enumerate}\setlength\itemsep{2pt}
\item Construct a realistic pseudo-data generator with an analytic, normalized
      ground-truth profile $s_{\rm true}(x,k,Q)$ and the exact forward map to $S$.
\item Run the full pipeline on noiseless pseudo-data (closure) and compute
      $\mathcal{A}[s]$.
\item Fit the real data once to obtain replica means for $S$ and $s$.
\item Use these means as the \emph{truth} for a second pseudo-data round to
      \emph{tune} evaluator settings (radial/azimuthal quadrature orders,
      adaptive-node parameter, taper onset, tail mask) and anchor weights
      $(\lambda_{\rm CS},\lambda_{m_2},\lambda_{\rm curv},\lambda_{\rm tail},\ldots)$,
      reducing residual structure.
\item Iterate steps 2--4 until $\mathcal{A}[s]$ and per-panel residuals plateau.
\end{enumerate}
The resulting noiseless bias $\Delta_{\rm acc}(x,k,Q)$ is assigned to the Stage-II
methodological \emph{accuracy}. Varying the pseudo-generator (e.g.\ smooth
deformations of widths/tails within a narrow neighborhood of the calibrated
truth) yields a spread of $\mathcal{A}[s]$ that we report as the Stage-II
methodological \emph{precision}. By construction, these methodological components
are \emph{not} produced by the MC replica loop; once their scales are fixed by the
recursion above, they are added a posteriori as a dedicated systematic.

\paragraph{Assumptions not included in the methodological error:}
Certain analysis choices are treated as fixed assumptions and are therefore
\emph{excluded} from the methodological variation accounting: (i) beam--target
interchange symmetry $S(q_T;x_1,x_2,Q)=S(q_T;x_2,x_1,Q)$; (ii) the positivity
constraint $s(x,k;Q)\ge 0$ (hence $f_{1,q/N}=f_q(x;Q^2)\,s \ge 0$); and (iii)
inverse-problem stabilizers (moment anchors, curvature regularization, tail
penalties, tapering), which improve conditioning at the cost of some
expressiveness. These ingredients are stress-tested in the methodological study,
but any residual loss of information is not fully quantifiable with the present
data alone.

\paragraph{Combination and intervals:}
Unless stated otherwise, central values are replica means and uncertainty bands
are empirical $68\%$ central intervals of the \emph{full} ensemble. Quantile
intervals are used for robustness against non-Gaussian tails. Methodological
terms are incorporated by augmenting the ensemble with the calibrated systematic
described above.

\begin{table*}[t]
\centering
\caption{Phase-space--averaged uncertainties (relative to the TMD central value).
Values are global averages across the resolved $(x,k_\perp,Q)$ region.
All sources are propagated through the MC ensemble except the methodological terms,
which are determined recursively (closure/tuning) and added a posteriori.}
\label{tab:uncertainty_budget}
\setlength{\tabcolsep}{8pt}
\renewcommand{\arraystretch}{1.15}
\begin{tabular*}{\textwidth}{@{\extracolsep{\fill}} l c}
\hline
\textbf{Source} & \textbf{Average scale and comment} \\
\hline
Experimental
&$\approx12\%$ stat. + $25\%$ ($15\%$) norm. E288 (E605); + $10\%$ syst. (E605)\\[0.25em]

Collinear PDF inputs
& $4$--$9\%$ (typ.\ $6$--$7\%$); PDF replicas \\[0.25em]

Algorithmic / numerical
& $1$--$3\%$ (typ.\ $2\%$); DNN retraining + evaluator scans \\[0.25em]

Methodological
& $<1.5\%$ total; inverse reconstruction of $s(x,k_\perp,Q)$ \\[0.15em]
\quad Accuracy component
& $\sim1.0\%$ phase-space--averaged residual in $s$. \\[0.15em]
\quad Precision component
& $\sim0.2\%$ generator-variation width; other method terms $<0.3\%$. \\
\hline
\end{tabular*}
\end{table*}

The methodological accuracy is set by the phase-space--averaged residual of $s$
in the recursive closure study; additional methodological contributions are found
to be smaller and are included in the quoted precision line. The numerical values
in Table~\ref{tab:uncertainty_budget} are intended as summaries; per-bin bands
always reflect the full MC ensemble (augmented by the methodological systematic)
at the quoted $68\%$\,CL.

\subsection*{A comparison with other work}

\begin{figure}[ht!]
  \centering
\includegraphics[width=0.45\textwidth]{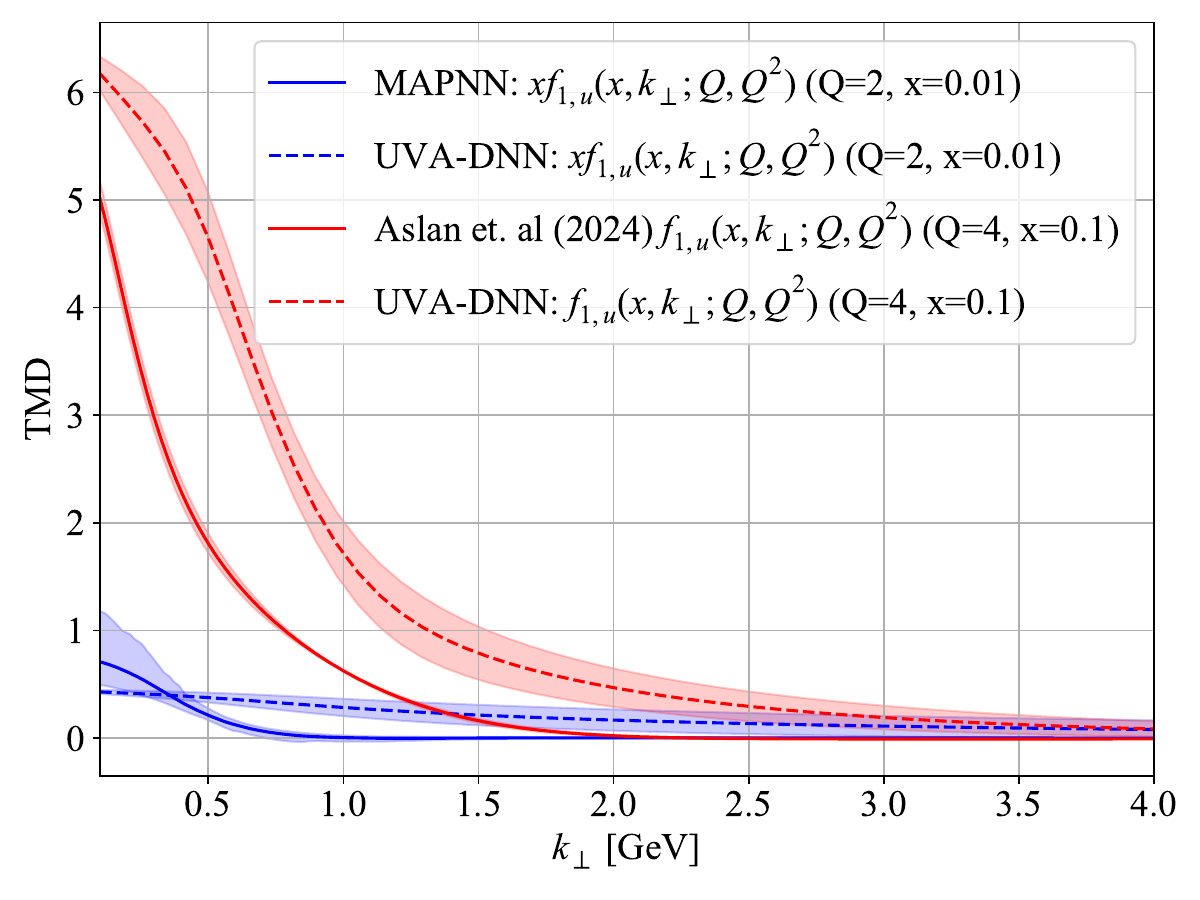}
\caption{Comparison of unpolarized $u$-quark TMD distributions.
Results from \textcolor{black}{MAPNN \cite{Bacchetta2025} ({\it solid-blue-curve})}, obtained from a global analysis including
fixed-target and hadronic collider data (Tevatron, RHIC, and LHC) with higher-order
perturbative ingredients; from \textcolor{black}{Aslan et al. (2024) \cite{Aslan_Ted_2025} ({\it solid-red-curve})}, based on a fit to E288
fixed-target data; and from this work \textcolor{black}{({\it dashed-curves})}, constrained by E288 and E605 fixed-target
Drell--Yan data and performed with a tree-level Drell--Yan hard factor and no
imposed Sudakov kernel or $Y$ term.
The comparison is intended as a qualitative illustration of relative shape and
magnitude rather than a matched perturbative-accuracy test.
}
  \label{fig:comparison}
\end{figure}

A qualitative comparison of $x f_{1,u}(x,k_\perp;Q,Q^2)$ with the results presented in
Ref.~\cite{Bacchetta2025}, and of $f_{1,u}(x,k_\perp;Q,Q^2)$ with those reported in
Ref.~\cite{Aslan_Ted_2025}, is shown in Fig.~\ref{fig:comparison}.
This comparison is intended for orientation only and should not be interpreted as a
like-for-like comparison across extractions with different perturbative content and
datasets.
In particular, the results of Ref.~\cite{Bacchetta2025} are obtained from a global analysis
including fixed-target data from Fermilab together with hadronic collider data from the
Tevatron, RHIC, and the LHC, and incorporate higher-order perturbative ingredients,
whereas Ref.~\cite{Aslan_Ted_2025} is based on a fit to E288 data alone.
By contrast, the present extraction is constrained only by fixed-target Drell--Yan data
from E288 and E605 and is intensionally performed with a tree-level Drell--Yan hard factor and without
imposing an explicit resummed Sudakov kernel or a fixed-order $Y$ term.  In addition, the analyses vary in how comprehensive uncertainty quantification and error propagation is handled.
For convenience, both TMD representations, $x f_{1,u}(x,k_\perp;Q,Q^2)$ and
$f_{1,u}(x,k_\perp;Q,Q^2)$, are displayed in the same figure.

\section{Conclusion}\label{sec:conclusion}

We have presented a momentum--space, physics-informed deep-learning framework that extracts purely data-driven
unpolarized TMDs directly from fixed-target Drell--Yan data without mapping to impact-parameter
space for fitting. The method embeds a normalized, nonparametric profile $s_h(x,k_\perp;Q)$ as
the \emph{integrand} of the $k_\perp$ convolution and trains it end-to-end against cross sections
through a differentiable quadrature layer. A two-stage pipeline first learns the structure kernel
$S(q_T;x_1,x_2,Q_M)$ at the cross-section level and then solves the inverse problem to reconstruct
$s_h(x,k_\perp;Q)$, enforcing per-$(x,Q)$ normalization.
This construction captures the joint $(x,k_\perp,Q)$ correlations supported by the data while
avoiding restrictive factorized functional Ans\"atze in the transverse sector.

Applied to E288/E605, the approach reproduces the measured $q_T$ spectra panel-by-panel in the
small-$q_T$ domain, demonstrating that the learned integrand and forward map are well conditioned
within the kinematic support of the data (Figs.~\ref{fig:cs_result_288} and \ref{fig:cs_result_605}).
The extracted TMDs exhibit the expected flavor hierarchy and smooth profiles in
$k_\perp$ that respect per-$(x,Q)$ normalization. Global surfaces in $(x,k_\perp)$ at fixed $Q$
(Figs.~\ref{fig:u3D_Q45} and \ref{fig:u3D_Q75}) and complementary $(x,Q^2)$ slices at fixed
$k_\perp$ (Fig.~9) make the learned multidimensional structure explicit. Over the fixed-target
hard-scale window, the extracted scale dependence is mild but systematic, consistent with the
limited lever arm in $Q_M$ of these data; in the present implementation it should be interpreted
as an effective, empirically learned $Q$ dependence within the fitted small-$q_T$ domain.

To characterize the transverse width through its second moment,
we introduce the RMS width
$\sigma_k(x,Q_M)=\sqrt{\langle k_\perp^2\rangle-\langle k_\perp\rangle^2}$ computed from normalized moments of the reconstructed distributions. The resulting field $\sigma_k(x,Q_M)$ exhibits a pronounced
dependence on $x$ and only a moderate variation with $Q_M$ (Fig.~\ref{fig:u-rms}) particularly at high $x$, providing a compact
summary of the transverse width that is insensitive to overall collinear normalization and thus
isolates shape information learned in momentum space.

Uncertainties are propagated through Monte-Carlo replica ensembles (1000 replicas) and reported as
68\% central intervals. We distinguish experimental (cross-section) and collinear (PDF) inputs
from algorithmic components, and we include a separate methodological term associated with the small systematic deviations through end-to-end closure tests. The latter is quantified using a recursive pseudo-data/closure schema performed on a tuned pseudo-data generator, followed by iterative refits to the experimental cross section in which
replica means from previous passes define an effective ``truth'' until residual deviations
stabilize. The resulting phase-space--averaged residual in $s_h$ defines a methodological accuracy
component that is added a posteriori as a dedicated systematic. The final phase-space--averaged budget (Table~\ref{tab:uncertainty_budget}) shows
experimental and PDF contributions as leading terms, with subdominant algorithmic uncertainty and a
small but explicitly quantified methodological accuracy from the DNN fit to the cross section and the inverse reconstruction.

The present implementation is intentionally designed to be maximally data-driven in the transverse
sector. We restrict to the TMD region $q_T<0.2\,Q$, set $\mu=\sqrt{\zeta}=Q$, employ a tree-level
hard factor, and do not impose an explicit resummed Sudakov kernel, fixed-order matching, or a
$Y$ term; consequently no $\mathrm{N}^k\mathrm{LL}$ resummation label is assigned. In addition,
the fixed-target lever arm in $Q_M$ bounds sensitivity to scale dependence, and the absence of
published covariance matrices requires diagonal replica resampling of the reported uncertainties.
Finally, in the present DY-only fixed-target study we adopt a flavor-independent transverse profile
within each hadron, with flavor dependence entering through the collinear PDFs; broader datasets
are required to robustly constrain fully flavor-dependent intrinsic $k_\perp$ shapes.

These scope choices do not limit the generality of the inverse-problem framework itself. The
central methodological advance is the differentiable inverse-operator strategy---neural integrand,
physics constraints, and end-to-end training through the $k_\perp$ convolution---which provides a
transferable template for extracting other intrinsic partonic distributions defined implicitly by
multidimensional integral transforms. The same machinery can be extended modularly to incorporate
higher-order perturbative ingredients (hard factors, explicit evolution kernels, matching and
$Y$ terms when desired) if one wishes to enforce a chosen perturbative accuracy, and the strategy can be
implemented equivalently in $k_\perp$ or $b_T$ space. Likewise, the framework generalizes
straightforwardly to flavor-dependent profiles $s_{q/h}$, hadron- and nucleus-dependent transverse
structure, and global datasets including collider DY/$Z$ and SIDIS that enlarge the lever arms in
both $x$ and $Q$. In this way, physics-informed differentiable inversion provides a scalable,
low-bias methodology for future extractions that remain close to measured spectra while enabling
systematic uncertainty accounting within a unified Monte-Carlo workflow.

\begin{acknowledgments}
The authors would like to thank Michael Wagman and Ted Rogers for their insightful discussions.We also thank Ted Rogers, J. O. Gonzalez-Hernandez (HSO-Approach), and C. Bissolotti (MAP Collaboration) for providing their TMD data points for the comparison. 
The authors acknowledge Research Computing at the University of Virginia for providing
computational resources and technical support that have contributed to the results reported in this publication.
For additional information, see \url{https://rc.virginia.edu}.
This work was supported by the U.S. Department of Energy under contract DE-FG02-96ER40950.
\end{acknowledgments}

\bibliography{Refs-update}

\end{document}